\documentclass[fleqn,usenatbib]{mnras}

\usepackage[T1]{fontenc}

\DeclareRobustCommand{\VAN}[3]{#2}
\let\VANthebibliography\thebibliography
\def\thebibliography{\DeclareRobustCommand{\VAN}[3]{##3}\VANthebibliography}

\usepackage{graphicx}	
\usepackage{amsmath}	
\usepackage{amssymb}	
\usepackage{multirow}
\usepackage{cuted}
\usepackage{hyperref}

\usepackage[dvipsnames]{xcolor}
\usepackage{soul}
\usepackage{newtxtext,newtxmath}

\title[Close white dwarf binaries~I: Reference samples]{Towards a volumetric census of close white dwarf binaries~I.\\
Reference samples}

\author[K. Inight et al.]{
K. Inight,$^{1}$\thanks{E-mail: u1983959@live.warwick.ac.uk}
Boris T. G\"ansicke,$^{1}$
E. Breedt,$^{2}$
T. R. Marsh,$^{1}$
A. F. Pala$^{3}$
and R. Raddi$^{4}$ 
\\
$^{1}$Department of Physics, University of Warwick, Coventry, CV4 7AL, UK\\
$^{2}$Institute of Astronomy, University of Cambridge, Cambridge, CB3 0HA, UK\\
$^{3}$European Southern Observatory, Karl Schwarzschild Straße 2, Garching, 85748, Germany\\
$^{4}$Departament de F\'isica, Universitat Polit\'ecnica de Catalunya, c/Esteve Terrades 5, E-08860 Castelldefels, Spain
}

\date{Accepted XXX. Received YYY; in original form ZZZ}

\pubyear{2020}

\begin{document}
\label{firstpage}
\pagerange{\pageref{firstpage}--\pageref{lastpage}}
\maketitle

\begin{abstract}
Close white dwarf binaries play an important role across a range of astrophysics, including thermonuclear supernovae, the Galactic low-frequency gravitational wave signal, and the chemical evolution of the Galaxy. Progress in developing a detailed understanding of the complex, multi-threaded evolutionary pathways of these systems is limited by the lack of statistically sound observational constraints on the relative fractions of various sub-populations, and their physical properties. The available samples are small, heterogeneous, and subject to a multitude of observational biases. Our overarching goal is to establish a volume-limited sample of all types of white dwarf binaries that is  representative of the underlying population as well as sufficiently large to serve as a benchmark for future binary population models. In this first paper, we provide an overview of the project, and assemble reference samples within a distance limit of 300\,pc of known white dwarf binaries spanning the most common sub-classes: post-common envelope binaries containing a white dwarf plus a main sequence star, cataclysmic variables and double-degenerate binaries. We carefully vet the members of these ``Gold'' Samples, which span most of the evolutionary parameter space of close white dwarf binary evolution. We also explore the differences between magnitude and volume limited close white dwarf binary samples, and discuss how these systems evolve in their observational properties across the \textit{Gaia} Hertzsprung-Russell diagram. 
\end{abstract}




\section{Introduction}
Binaries are important in many aspects of astrophysics such as their role as precursors of supernovae \citep{2018PhR...736....1L,2012Natur.481..164S}, detectable sources of gravitational wave radiation \citep{2018arXiv180701060N} and in the chemical evolution of galaxies \citep{1995MNRAS.277..945T,2004NewAR..48..861D}. Binary evolution involves complex physical processes including common envelope evolution and orbital angular momentum loss that are not well understood \citep{2013A&ARv..21...59I,2011ApJS..194...28K}. Models describing these physical processes involve a number of free parameters, that require observational calibrations. However, the currently known samples of different types of binaries are severely incomplete, and subject to observational biases that are difficult to quantify \citep{2007MNRAS.374.1495P,2004RMxAC..20..152G}. Consequently, predictions of binary populations have to be considered with some reservations, as the parameters used to encapsulate the physics of binary evolution are not well calibrated. 

Progress requires the identification of representative populations of binaries with well-understood selection effects that sample the entire phase-space of physical parameters, as well as evolutionary stages. The key motivation of the project presented here is to assemble a volume-limited sample that includes a homogeneous representation of all major sub-types of close white dwarf binaries: detached white dwarf plus main-sequence post-common envelope binaries (PCEBs), interacting white dwarf binaries (cataclysmic variables,  CVs), and double white dwarfs (DWDs).  One example of such a study is the joint population study of all types of white dwarfs, single and in binaries, within the volume-limited and complete 20\,pc sample \citep{2017A&A...602A..16T}.

The key information required for constructing a truly representative sample of white dwarf binaries is the accurate knowledge of their distances. \textit{Hipparcos} measured parallaxes for $\simeq118\,000$ stars brighter than $V\simeq11$ \citep{1997ESASP1200.....E}, and included only a small number of CVs \citep{1999IBVS.4731....1D}, as most of these systems are intrinsically faint. Ground-based parallax measurements for a few dozen CVs extended to fainter magnitudes, but were limited in accuracy \citep{2008AJ....136.2107T, 2003AJ....126.3017T}. Precise space-based parallaxes were obtained for a handful of selected CVs using the \textit{Hubble Space Telescope} (\textit{HST}) \citep[e.g.][]{2000AJ....120.2649H, 2003A&A...412..821B, 2004A&A...419..291B, 2007ApJ...666.1174R}, although, some of those measurements sparked controversies \citep{2002A&A...382..124S, 2004AJ....127..460H, 2013Sci...340..950M}. Accurate distances to faint stars have only now become systematically available, thanks to the all-sky astrometry provided by the \textit{Gaia} mission. In Data Release~2\footnote{Data Release~1 only included parallaxes of $\simeq2$ million stars from a joint \textit{Tycho}-\textit{Gaia} astrometric solution, including 16 cataclysmic variables \citep{2017A&A...604A.107R}.} (DR2, \citealt{2018A&A...616A...1G}) \textit{Gaia} provides parallax measurements for $\sim$1.3 billion stars, with a limiting magnitude of $G\simeq20$. 

\citet{2020MNRAS.494.3799P} used the \textit{Gaia} DR2 data to establish the first truly volume-limited study of CVs within 150\,pc. Whilst limited to only 42 systems this study provided the most robust estimate of the space density of CVs to date  ($\rho_0=4.8\genfrac{}{}{0pt}{}{+0.6}{-0.8}\times10^{-6}\,\mathrm{pc^{-3}}$), as well as an assessment of the intrinsic make-up of of the CV population, with the surprising conclusion that about one third of all CVs contain a magnetic white dwarf. 

With the \textit{Gaia} astrometry of $\simeq1.3$~billion stars in hand, the limitation in extending the study of \citet{2020MNRAS.494.3799P} to a larger volume, and including all types of close white dwarf binaries, is the incompleteness of the known members of the different sub-classes. \citet{2020MNRAS.494.3799P} estimated a $\simeq77$\,per cent completeness for the CVs within 150\,pc, that value drops for larger distances, and is poorly constrained for PCEBs.  

The first step within this project is to establish reference samples of each sub-type of white dwarf binaries that are as representative as possible. The definition of these samples, and a discussion of their overall properties within the \textit{Gaia} Hertzsprung-Russell (HR) diagram is the focus of this paper. These reference samples will then be used to define algorithms that identify white dwarf binary candidates, combining \textit{Gaia} data with observations extending over a wider wavelength range. 

In this paper we identify the key characteristics of each close white dwarf binary sub-type (Sect.\,\ref{sec:subtypes}), discuss the limitations of \textit{Gaia} (Sect.\,\ref{sec:gaia}), define the reference ``Gold'' samples for each sub-type (Sect.\,\ref{sec:refsamples}) including a discussion of the selection effects that affect these samples, and finally use the Gold Samples to discuss the evolution of close white dwarf binaries in the \textit{Gaia} Hertzsprung-Russell diagram  (Sect.\,\ref{sec:evolution}).

\section{Close white dwarf binaries}\label{sec:subtypes}
\begin{figure*} 
\includegraphics[width=17cm]{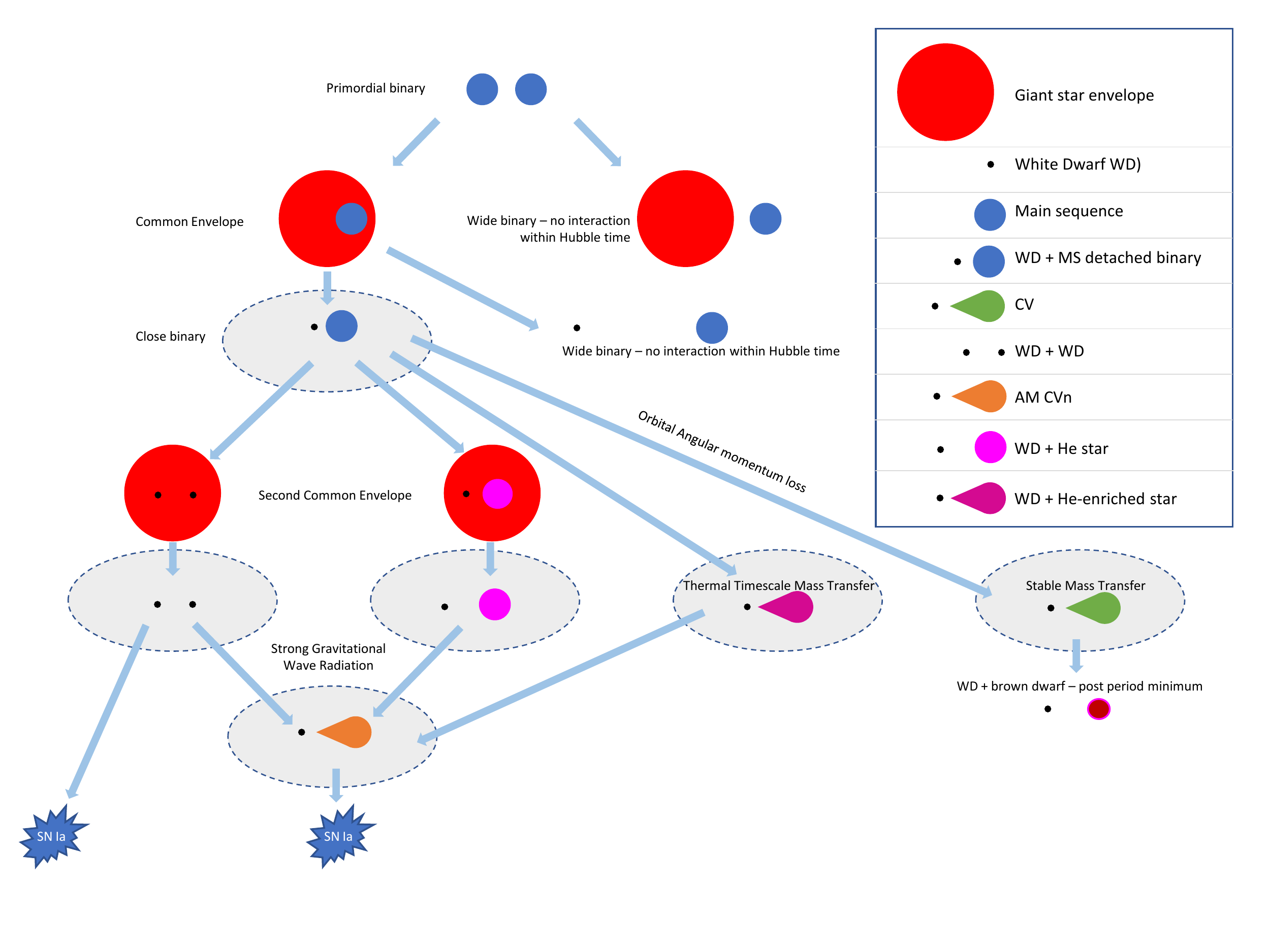}
\caption { A schematic overview of the evolutionary channels leading to the different sub-classes of close white dwarf binaries discussed in this paper. A number of these channels potentially lead to the ignition of white dwarfs as type Ia supernovae and to the emission of gravitational waves detectable by the \textit{LISA} mission, highlighting the importance of these systems for a wide range of astrophysical problems. The grey ellipses show the sub-classes that are included in our census. Adapted from   \protect\citet{2019BAAS...51c.168T} \label{fig:02} 
}
\end{figure*}

Within this project, we will focus on close white dwarf binaries that have  undergone interactions in the past, in the vast majority of cases in the form of a common envelope. Some of these are currently in a mass transferring state, others are in a detached configuration but may interact again in the future. The discussion in this and the following papers will frequently refer to these physical properties, as well as the past and future evolution of the major classes of close white dwarf binaries. A graphical overview of the evolutionary links between the different types of white dwarf binaries discussed below is shown in Fig.\,\ref{fig:02}.

\subsection{Post Common Envelope Binaries (PCEBs)}
\label{sec:PCEBsIntro}

Between 50 and 80 \,per cent of stars are formed as part of a binary system \citep{2008MNRAS.389..869E,2010ApJS..190....1R}. Both stars will evolve normally through the main sequence until the higher mass primary runs out of hydrogen in its core and ascends the giant branch. If the initial binary separation is sufficiently small, it will eventually fill its Roche lobe, initiating unstable mass transfer onto the lower-mass companion \citep{1976IAUS...73...75P,  1984ApJ...277..355W}. The mass transfer timescale is faster than the rate that the companion can adjust to and the two stars are rapidly engulfed in a common envelope. Friction within this envelope drains energy and  angular momentum from the binary orbit, shrinking their separation. The envelope acquires this angular momentum and is ejected leaving a binary consisting of a white dwarf and a main-sequence star. 

The orbital shrinkage caused by common envelope evolution implies that the resulting population of white dwarf binaries is bi-modal (see Figure~10 in \citealt{2004A&A...419.1057W}) with post common envelope binaries (PCEBs) having periods of hours to days, and wide white dwarf plus main sequence star binaries (that avoided a common envelope and evolved as single stars) having periods in excess of a few years. We note that depending on the details of the initial binary parameters,  a small number of PCEBs can emerge with periods of $\sim100$\,d \citep{2014Sci...344..275K, 2014A&A...568L...9Z}. This bi-model distribution has been observationally confirmed with high-resolution \textit{HST} imaging and radial velocity studies of a sample of white dwarf binaries \citep{2010ApJS..190..275F, 2019MNRAS.484.5362A}.

Identifying PCEBs is subject to two major selection effects: on the one hand, PCEBs in which one component dominates the observed wavelength range will resemble either single white dwarfs, or main-sequence stars. On the other hand, with typical distances of a few 100\,pc, PCEBs will be indistinguishable from spatially unresolved wide binaries separated by a few to a few tens of au, even at \textit{Gaia}'s space-based resolution. These selection effects will be discussed in more detail in Section\,\ref{subsec:WD+m}.

\subsection{Cataclysmic Variables (CVs)}
\label{sec:CVsIntro}
Once orbital angular momentum loss reduces the orbital separation of PCEBs such that the main-sequence star fills its Roche lobe, mass transfer onto the white dwarf is initiated via the inner Lagrangian point between the two stars; the systems morph into cataclysmic variables (CVs), with observational properties that radically differ from those of their detached progenitors. 

Most CVs above the period gap have periods $\la14$\,h, with a small number of systems with donors that have slightly evolved off the main-sequence have periods extending up to a few days \citep{1992Natur.358..563S}. As a CV evolves and the companion loses mass, the separation reduces and the orbital period decreases. The observed distribution of orbital periods shows a significant dip between $2-3$\,h \citep{1980MNRAS.190..801W}, and this deficiency is known as the \textit{period gap}. The currently accepted explanation for the period gap is that CVs undergo a major drop in the rate at which they lose angular momentum once the donor stars become fully convective \citep{1983ApJ...275..713R, 2001ApJ...550..897H}, at $\simeq3$\,h, and evolve as \textit{detached} systems across the period gap \citep{2008MNRAS.389.1563D, 2016MNRAS.457.3867Z}, until the Roche-lobe shrinks sufficiently to, once again, establish a semi-detached configuration, re-starting the mass transfer, albeit at a much lower rate. As mass transfer keeps eroding the mass of the donor star, it eventually becomes a brown dwarf. The result of the change in the internal structure of the donors is that further evolution will drive them back towards longer periods \citep{1983ApJ...268..825P}, and hence there is minimum orbital period, $P_\mathrm{min}$. As CVs evolve twice through the period range near $P_\mathrm{min}$, there is a pile-up of systems which has been observationally confirmed with $P_\mathrm{min}\simeq80$\,min \citep{2009MNRAS.397.2170G}.

\subsubsection{CV sub-types}
Historically, a large number of CV sub-classes have been defined based on these  observational characteristics, which, ideally, should map onto a small number of fundamental physical properties of these binaries. Three such defining key parameters are the orbital period ($P_\mathrm{orb}$) of the system, magnetic field strength ($B$) of the white dwarf, and the mass loss rate ($\dot M$) of the companion (or donor) star. These various sub-classes of CVs with different accretion geometries present distinctly different spectroscopic appearances, which will affect our ability to identify them (Sect.\,\ref{sec:CVs}).

In the absence of a strong magnetic field, an accretion disc forms around the white dwarf. These discs can be subject to a limit cycle based on a thermal instability \citep{1981A&A...104L..10M}, in which they undergo quasi-periodic large changes in  disc temperature and brightness~--~observationally CVs that exhibit these disc outbursts are called dwarf novae. Whether or not an accretion disc is subject to this instability depends on both the size of the disc (and hence on the orbital separation, and in turn on $P_\mathrm{orb}$) and on $\dot M$ \citep{1986ApJ...305..261S, 1992ApJ...394..268S}. If the disc is sufficiently small, and $\dot M$ sufficiently high, the disc will be in a steady, hot, bright state (akin to a dwarf nova permanently in outburst), and such CVs make up the sub-class of nova-like variables, or novalikes for short. 

Dwarf novae are the dominant CV sub-type in a volume-limited sample \citep{2020MNRAS.494.3799P}, and are further divided according to their outburst behaviour. SU\,UMa systems   \citep{2020PASJ...72...14K} exhibit, in addition to the normal short outbursts lasting a few days, superoutbursts that are somewhat brighter, and can last several weeks. These superoutbursts are triggered when the outer radius of the disc reaches a 3:1 resonance, resulting in tidal interactions with the companion star \citep{1989PASJ...41.1005O}. These interactions force the accretion disc to precess, resulting in a brightness modulation, so-called \textit{superhumps}, on the beat period between the precession period of the disc and the orbital period~--~i.e. somewhat longer than the orbital period \citep{2005PASP..117.1204P}. The majority of the SU\,UMa systems are found below the period gap. WZ\,Sge systems are a sub-class of dwarf novae defined by having extremely long outburst recurrence times of years to decades \citep{2002PASP..114..721P, 2015PASJ...67..108K}, implying that they have low mass transfer rates, and most of them have orbital periods in the range $\simeq75-95$\,min. Dwarf novae that \textit{do not} undergo superoutbursts are almost exclusively found above the period gap, and further divide into  U\,Gem systems \citep{1984PASP...96....5S} that exhibit quasi-periodic outbursts, and Z\,Cam systems \citep{1983A&A...121...29M} that alternate between states where they undergo disc instabilities, and so-called stand-stills during which the disc remains in a hot, stable state, akin to the novalike variables. 

If the magnetic field of the white dwarf is sufficiently strong ($B\ga10$\,MG) to control the accretion geometry, essentially such that the Alfv\'en radius of the white dwarf is  larger than its Roche lobe, the formation of an accretion disc is suppressed. The material lost by the donor initially follows  a ballistic trajectory, and is then channelled along the magnetic field lines onto the magnetic poles of the white dwarf. These CVs go by the name \textit{AM\,Her systems} (after their prototype, \citealt{1977ApJ...213L..13C}) or \textit{polars} (because of the strongly polarised cyclotron radiation they emit, \citealt{1977ApJ...212L.125T}). Interactions of the strong magnetic field of the white dwarf with that of the secondary star  lock the white dwarf into synchronous rotation with the orbital period \citep{1983ApJ...274L..71L}. An early overview of the study of polars is given by \citet{1990SSRv...54..195C} 

In the case of an intermediate magnetic field strength ($1\la B\la10$\,MG), a vestigial disc may form, disrupted at the Alfv\'en radius, from where the material is again funnelled onto the magnetic poles of the white dwarf. The magnetic field strengths in these \textit{intermediate polars} (sometimes called \textit{DQ\,Her stars}), are poorly known, as usually neither Zeeman-split photospheric absorption lines from the photosphere, nor cyclotron emission from the accretion funnel, are observed. The white dwarf spin periods in intermediate polars are always shorter than the orbital periods, and some of them may evolve into polars as their period shortens and mass transfer decreases \citep{2004ApJ...614..349N, 2008ApJ...672..524N}. An overview of the properties of intermediate polars is given by \citet{1994PASP..106..209P}.

A sub-class of CVs with distinct physical properties, and evolutionary histories, are the AM\,CVn stars: binaries containing a white dwarf accreting from a hydrogen-deficient donor star that is fully or partially degenerate \citep{2005ApJ...624..934D}, with orbital periods of $\simeq5-68$\,min \citep{2015MNRAS.446..391L, 2018A&A...620A.141R, 2020MNRAS.496.1243G}. These systems are important in the context of gravitational wave physics, as they will be easily detected by the \textit{LISA} space mission \citep{2018MNRAS.480..302K}. 

Finally, there are a handful of exotic, and nearby, white dwarf binaries such as the radio-pulsating AR\,Sco \citep{2016Natur.537..374M} that fall outside the current evolutionary models.

\subsection{Double White Dwarfs (DWDs)}
Double white dwarfs (DWDs) \citep{1995MNRAS.275..828M} are often  formed as the result of a second common envelope event. Others may have formed the first white dwarf in a period of stable mass transfer (Algol-like) to the companion. DWDs lose angular momentum through gravitational radiation and their separation will decrease~--~eventually leading to a merger. If their combined mass is sufficiently large, this merger may result in a thermonuclear supernova \citep{2012A&A...546A..70T, 2018ApJ...854...52S}. DWDs are the most common, and have the best-determined space density, of all close white dwarf binaries, thanks to the detailed studies of the local 20\,pc sample: \citet{2017A&A...602A..16T} lists one confirmed, and six candidate DWDs (see their table\,1) within 20\,pc. Four of those candidates were confirmed as DWDs based on their \textit{Gaia} parallaxes, with two remaining possible candidates \citep{2018MNRAS.480.3942H}, implying that $\simeq3-5$\,per cent of all apparently single (i.e. spatially unresolved) white dwarfs are actually DWDs, with a space density of $\simeq(1.4-2.3)\times10^{-4}\,\mathrm{pc^{-3}}$. Most of these DWDs are cool and would be very hard to identify at larger distances. A complete volumetric sample therefore will need to encompass a much smaller volume than that of CVs and PCEBs.

\section{From parallaxes to distances}\label{sec:gaia}
\begin{figure}
     \centering
    \begin{minipage}{0.47\textwidth}
    \centering
    \includegraphics[width=0.89\textwidth]{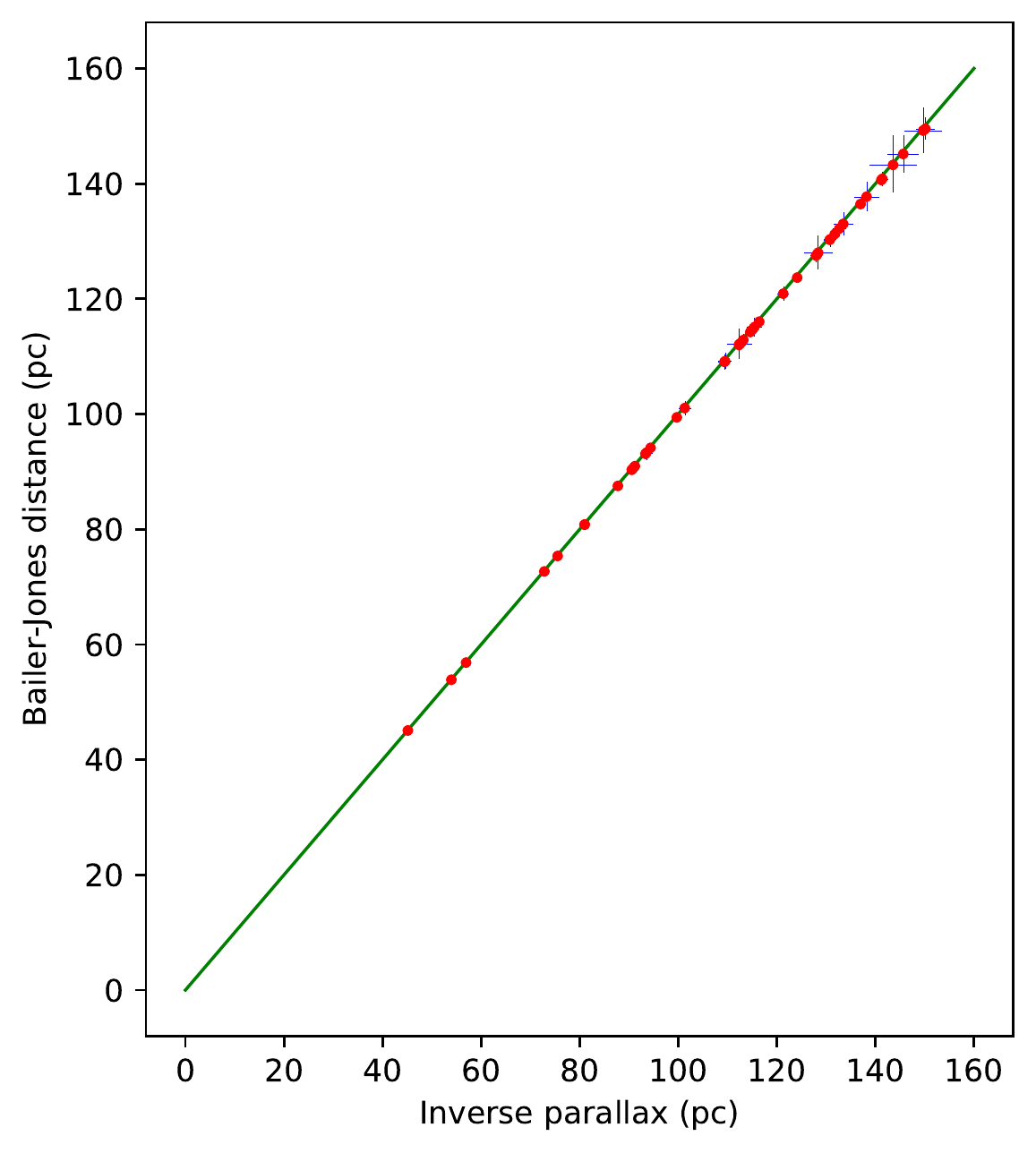}
    \end{minipage}
    \begin{minipage}{0.47\textwidth}
        \centering
        \includegraphics[width=0.89\textwidth]{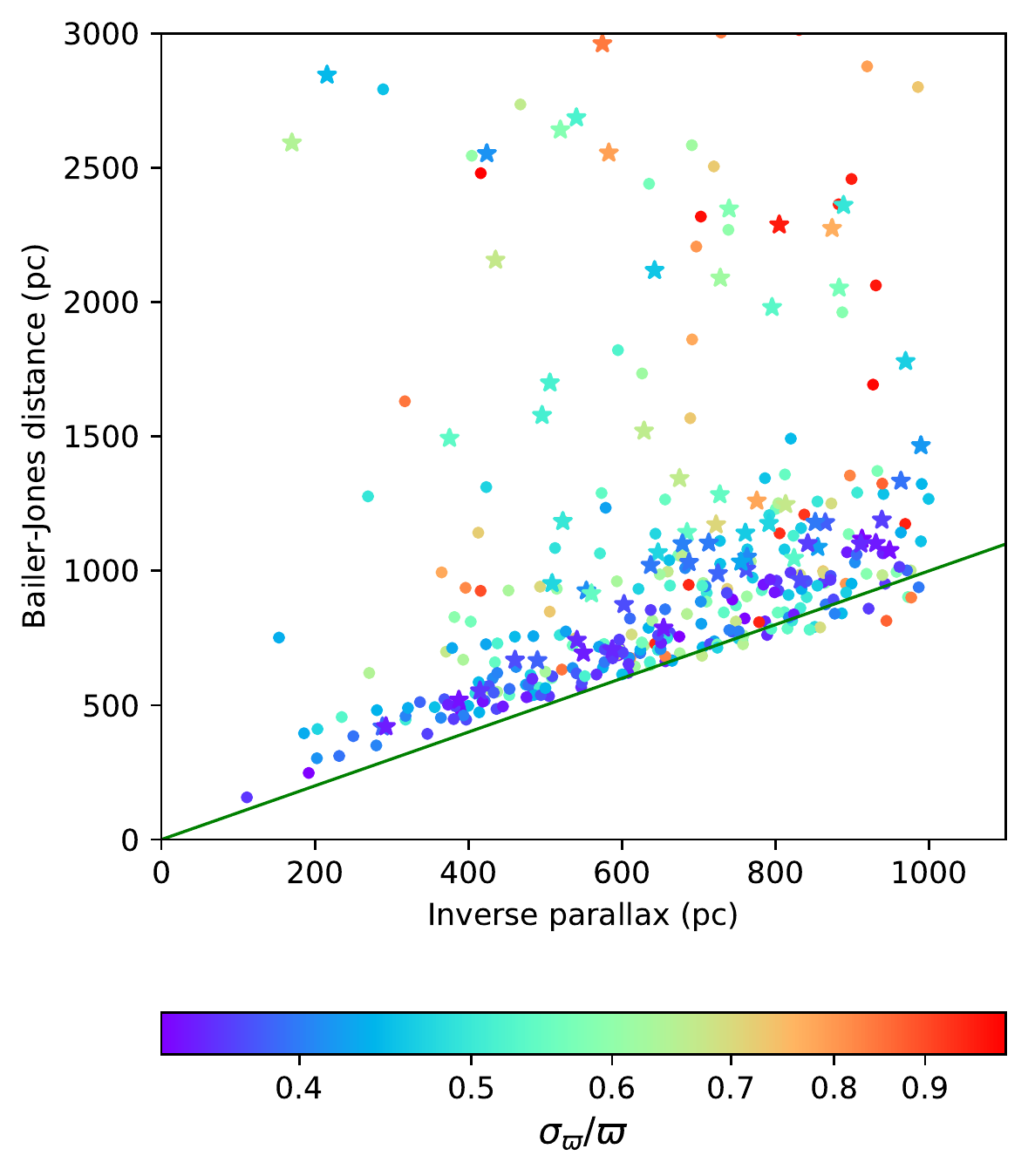}
    \end{minipage}

\caption{Comparison between $d=1/\varpi$ and the distances of \citet{2018AJ....156...58B}. Top panel: 42 CVs known within 150\,pc \protect\citep{2020MNRAS.494.3799P}, with $\sigma_\varpi \le 0.24$\,mas and $\varpi/\sigma_\varpi>27.7$, illustrating that for accurate parallaxes, the inverse of the parallax is a very good approximation for the distance. Bottom panel: 374 CVs and CV candidates with $\sigma_\varpi> 0.33$mas and $ 1 < \varpi/\sigma_\varpi< 3$. A star indicates that the object is outside the Galactic plane ($\lvert b \rvert>10^{\circ}$).  It is evident that for large parallax uncertainties, $1/\varpi$ often significantly underestimates the likely distance of the systems. In both panels the solid green line indicates a 1:1 correspondence.  
\label{fig:01}}
\end{figure}

The astrometry of the \textit{Gaia} space mission \citep{2016A&A...595A...1G, 2018A&A...616A...1G} is revolutionising Galactic and stellar astronomy. \textit{Gaia} has vastly increased the number of stars with accurate parallaxes,  $\varpi$,
however, a large fraction of the objects within DR2 have substantial parallax uncertainties, such that $d=1/\varpi$ is no longer a reliable estimate of their distances. \citet{2015PASP..127..994B}
has used a probabilistic model using the known geometric distributions of stars to derive distance estimates that avoid the errors in traditional methods, such as spectro-photometric estimates which are subject to assumptions regarding interstellar extinction and stellar effective temperatures. 
We illustrate (Figure~\ref{fig:01}) the difference between $d=1/\varpi$ and the distances of \citet{2018AJ....156...58B} using the sample of 42 CVs (top panel) known within 150\,pc \citep{2020MNRAS.494.3799P}, all of which have $\sigma_\varpi\le 0.24$\,mas and $\varpi/\sigma_\varpi>27.7$. This contrasts with the 374 CVs and CV candidates with distances beyond 1000\,pc  with $\varpi>1$\,mas and  $1 < \varpi/\sigma_\varpi< 3$ .  Figure~\ref{fig:01} demonstrates that $1/\varpi$ results in substantial underestimates with respect to the distances of \citet{2018AJ....156...58B} for data with large parallax uncertainties. We note from \citet{2018A&A...616A...2L} that faint sources or those in crowded areas of the sky are most susceptible to errors in the five-parameter solution  (and the parallax in particular). We have split the sources in Fig.\,\ref{fig:01} according to their location on sky, with those in crowded areas (defined as the galactic plane, $|b|<10^\circ$) shown as circles, and those in uncrowded areas as stars. It appears that for CVs and CV candidates, there is no clear correlation between the magnitude of the discrepancy between $1/\varpi$ and the distances of \citet{2018AJ....156...58B} and whether a source is found within a crowded or uncrowded region. 
\citet{2018AJ....156...58B} acknowledges that there are shortcomings in the relatively simple prior used and indeed has been developing more sophisticated models for Gaia EDR3 \citep{2020arXiv201205220B}. Nevertheless the \citet{2018AJ....156...58B} estimates (\texttt{r\_est}) are a significant improvement on simple inverse parallax estimates and we therefore have adopted them throughout this project.

One major shortcoming of \textit{Gaia} DR2 for the construction of volumetric samples is that 361\,009\,408 (21.3\,per cent)  of the objects in \textit{Gaia} DR2 do not have parallaxes and hence their distances are unknown. The most common reason is that those sources are too faint for \textit{Gaia} to obtain a parallax. Whilst this is not a major problem for the purpose of this paper (the definition of Gold Samples of various classes of white dwarf binaries) it is is a concern for the future papers of this project, as several sub-classes (such as low $\dot M$ CVs, DWDs, and binaries containing cool white dwarfs with very low-mass companions) are intrinsically faint. Hence, the incompleteness of \textit{Gaia} parallaxes is likely to reduce the size of our census, and may introduce biases. However, we expect the situation to improve with \textit{Gaia}~DR3. Here, we assess the significance of the lack of parallaxes of a significant fraction of the \textit{Gaia}~DR2 sources by establishing, for each white dwarf binary subclass individually, the fraction of previously known systems without a parallax that are likely to be within 300\,pc.

\section{Reference samples}\label{sec:refsamples}
The objective of this section is to define a prototypical Gold Sample for each type of close white dwarf binary based on the known systems that can then be used to define selection criteria for an all-sky search. The ultimate goal of this search is to establish a view of the Galactic white dwarf binary populations that is homogeneous, representative, and as complete as possible within a given volume limit. We emphasise that each Gold Sample will inevitably be subject to the selection effects underlying the known populations of white dwarf binaries. However, our goal here is not to achieve completeness at any level, but to map out the parameter space occupied by these white dwarf binaries.  

For the goal of completeness within a given volume, two aspects limit the maximum distance for which it will be possible to identify white dwarf binaries, i.e. the requirement for (1) a \textit{Gaia} parallax, and (2) spectroscopic confirmation. Within \textit{Gaia} DR2, parallaxes require a 5-parameter astrometric solution; which exists for $\simeq83$~per cent of targets with $G\le20$, dropping to $\simeq16$~per cent at $G\le21$ (interpolated from Figure B1 in \citealt{2018A&A...616A...2L}). The required follow-up spectroscopy of a large number ($\simeq100\,000$) of candidate white dwarf binaries can only be achieved via multi-object spectroscopic surveys, such as DESI \citep{2016arXiv161100036D, 2016arXiv161100037D}, WEAVE, \citep{2012SPIE.8446E..0PD}, 4MOST, \citep{2016SPIE.9908E..1OD}, and SDSS-V, \citep{2017arXiv171103234K}. These experiments will deliver spectroscopy with a sufficient quality for a reliable identification for sources as faint as $G\simeq20$

We therefore adopt a magnitude limit of $G=20$, and using  $m=20-5\log(d/10)$, we determine the intrinsically faintest systems (assuming minimal extinction) we can detect for a given distance limit, as illustrated in Figs.\,\ref{fig:09},\ref{fig:29},\ref{fig:33} for the different classes of white dwarf binaries. The overall volume limit of our census is a compromise between sample size and completeness and we have chosen 
\begin{align}
\texttt{r\_est}< \textrm{300 pc}\label{eq:300pc}
\end{align}
where \texttt{r\_est} is the distance estimate of \citet{2018AJ....156...58B}, with the exception of the DWDs (Sect.\,\ref{sec:ddgold}). This choice will introduce well-quantifiable selection effects (some types of white dwarf binaries are inherently fainter than others), which we explore further in the context of each individual sub-class. 

Accurate astrometry and photometry are needed to avoid misleading conclusions when we use the samples to evaluate potential cuts. Following \citet{2018A&A...616A...2L} and \citet{LL:LL-124} we filter the astrometry by: 
\begin{align}
&\texttt{ruwe}  < 1.4 \label{eq:1}\\
&\texttt{$\varpi/\sigma_\varpi $} > 10 \label{eq:2}
\end{align}
where \texttt{ruwe} is the ``re-normalised unit weight errors'', and $\varpi$ and $\sigma_\varpi$ are the parallaxes and its uncertainties.

To ensure that the precision of the red and blue photometry is within 10~per cent ($\simeq0.1$ mag)  we again followed  \citet{2018A&A...616A...2L} and filtered on:
\begin{align}
&\texttt{phot\_bp\_mean\_flux\_over\_error}  > 10\label{eq:3}\\
&\texttt{phot\_rp\_mean\_flux\_over\_error}  > 10\label{eq:4}
\end{align}
Known CVs that failed these quality cuts are listed in Table\,\ref{tab:app_CV_failed}. These filters do not just exclude dim stars. Two examples closer than 100\,pc are U\,Gem (excluded because $\texttt{ruwe}=1.48$) and AM\,Her (excluded because $\texttt{phot\_rp\_mean\_flux\_over\_error}=6.04$). Although it may seem counter-intuitive to omit well-known systems it must be remembered that  the purpose of a Gold Sample is not to be complete, but to be clean, with good \textit{Gaia} data. The significance of this will become apparent in the context of target selection in Paper~II.

The accurate astrometry enables the derivation of an important parameter of the various white dwarf binary classes: their space density.  The space density makes assumptions about the age and hence the scale height of systems and can be calculated using the approximation:-  
\begin{equation}
\rho=  \rho_0\,e^{-|z|/h} 
\end{equation}
where h is the scale height. The effective volume for a given radius ($R$) is then calculated by integration:-
\begin{equation}
V_\mathrm{eff} = {\int_{0}^{R} \int_{-\sqrt{R^2-x^2}}^{\sqrt{R^2-x^2}} \left(2\pi x\right) \left(\exp^{-|z|/h} \right)dz\, dx}\label{eq:scale_heights}
\end{equation}
Using this expression, the space density of the considered type of white dwarf binary is then given by the size of the sample divided by $\rho=N/V_\mathrm{eff}$. This requires an assumption for the value of $h$, which depends upon the age of the stars in the sample \citep{2007MNRAS.374.1495P}. Below, we follow \citet{2020MNRAS.494.3799P} and calculate the space densities of the Gold Samples for three different values of $h$ (100, 280, 500\,pc). The estimates of $\rho_0$ will represent lower limits to the true space densities of the respective white dwarf binary populations as the Gold Samples are not complete.


\begin{table*}
\caption{Assembling the Gold samples. Positive numbers indicate the initial sample size, negative numbers indicate the effects of the individual cuts that were subsequently applied. The requirement for a complete set of accurate attributes dramatically reduces the final sample size. $\ddag$ see Equations \ref{eq:1}-\ref{eq:4}} \label{tab:reconcilation} 

\setlength{\tabcolsep}{1.5ex}
\begin{tabular}{lrrrrrrrrrrrr}\hline
System   type                    & \multicolumn{5}{c}{WD+M}                               & \multicolumn{3}{c}{WD+AFGK}      & CV    & \multicolumn{3}{c}{DWD} \\ \hline
Source                           & SDSS  & Ashley~ & Zorotovic~ & RK~ &Other   & Ren~   & RK~    & Other     &       & 20\,pc & 40\,pc~  & Other  \\
                                 &       & sample  & sample     & sample && sample & sample &           &       & sample       & sample &        \\
Source Systems                   & 3287  & 16      & 25         & 190    &2& 23     & 25     & 7          & 5192  & 5            & 24     & 198    \\
Duplicates                       &   -1    & -6        & -3         & -131   & &        &        &    -2       &       &              & -3     & -2     \\
Systems with no \textit{Gaia} match       & -213      & -1      &            & -2     &&        &        &           &       &        &        &        \\
Systems with no parallax         & -429  &         &            & -3     &&        &        &           & -1150 &              &        & -6     \\
Systems $>$300pc                    & -2049 &       & -4         & -26   & & -10    & -23    &           & -3737 &              &        & -129    \\
Systems failing quality criteria $\ddag$ & -272    & -2      & -7         & -5     && -3     & -1     & -1        & -104   & -1    & -10    & -9     \\
Unreliable/invalid Systems       &      &         &          & -8     &&        &        &           & -50   &              &        &        \\
Wide binaries                    & -125  &         &            &        &&        &        &           &       &              &        &        \\
Possible wide binaries           & -119  &         &            &        &&        &        &           &       &              &        &        \\
   &        &           &       &              &        &        \\
Final Gold Sample                & 79    & 7      & 11         & 15     &2& 10     & 1      & 4         & 151   & 4            & 11     & 52     \\
           & \multicolumn{5}{c}{\upbracefill} &  \multicolumn{3}{c}{\upbracefill}&\multicolumn{1}{c}{\upbracefill}     & \multicolumn{3}{c}{\upbracefill} \\
          &\multicolumn{5}{c}{114}&\multicolumn{3}{c}{15}&151&\multicolumn{3}{c}{67} \\\hline
\end{tabular}
\end{table*}

\subsection{PCEBs: White dwarfs with M-type companions}\label{subsec:WD+m}
\subsubsection{Establishing the Gold Sample}
There has been a continuing effort over many years to build a comprehensive catalogue of binaries consisting of a white dwarf and a main sequence M-type star (hereafter WD+M, see \citealt{2007MNRAS.382.1377R, 2010MNRAS.402..620R, 2012MNRAS.423..320R, 2013MNRAS.433.3398R, 2016MNRAS.458.3808R}). This catalogue contains WD+M binaries identified from the detection of both stellar components in their SDSS spectroscopy. Whereas these systems are unresolved in the SDSS imaging data, given the typical spatial resolution $\simeq1.5$\,arcsec, and distances of several 100\,pc, they can be either short-period (hours to days) PCEBs or wide binaries with separations of tens of AU that never interacted. Extensive radial velocity studies of the SDSS WD+M binary sample showed that $\simeq21-24$~per cent are PCEBs, with a period distribution peaking at $\simeq10.3$\,h \citep{2011A&A...536A..43N}. In the online catalogue of SDSS WD+M binaries\footnote{\url{http://www.sdss-wdms.org}}, PCEBs are defined as systems where a $\ge3\sigma$ radial velocity variation has been detected in the spectroscopy, and we select these ``SDSS'' systems for our Gold Sample.

To these we added confirmed WD+M binaries from \citet{2006ApJ...646..480F, 2010ApJS..190..275F}, some of which were later followed up by \citet{2019MNRAS.484.5362A}.

These WD+Ms (hereafter the ``Ashley sample'') were identified from catalogues of white dwarfs due to their infrared excess and were unresolved in high-resolution \textit{HST} imaging. \citet{2019MNRAS.484.5362A} obtained radial velocity measurements of these systems, and corroborated the bi-modality of the orbital separations referred to earlier, i.e. that we can expect WD+M binaries to be either close (with periods of hours to days) or wide (with periods $>$ years). 

\begin{figure*} 
\includegraphics[width=\textwidth]{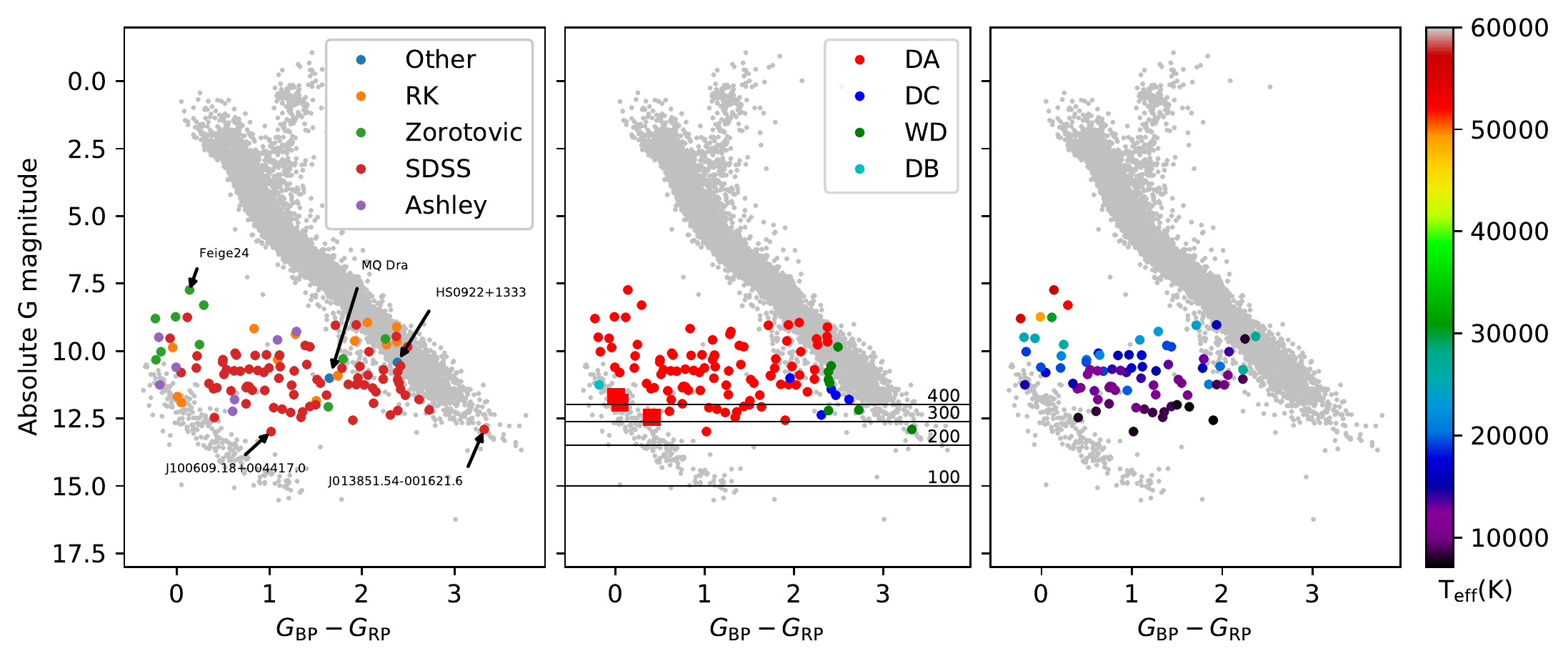}
\caption [.] {\label{fig:09} 
The WD+M Gold Sample in the HR diagram. Left panel: distribution of the sub-samples. Middle panel: distribution of the white dwarf spectral types. The concentration of systems with no spectroscopic classification (WD, green symbols) near the main sequence indicates that the companion dominates the optical flux of these systems. Systems with brown dwarf companions are indicated by squares; their location within the white dwarf cooling sequence is consistent with the low luminosity of the brown dwarfs. The minimum absolute magnitude to be detectable at an apparent magnitude of $G=20$ is indicated by grey lines for distances of 200, 300 and 400\,pc. Right panel: The white dwarf effective temperatures of the WD+M Gold Sample plotted on the HR diagram. The decrease in absolute magnitude with decreasing effective temperature is evident. }
\end{figure*}

We also added confirmed PCEBs from appendix A2 of  \citet{2010A&A...520A..86Z} (hereafter the ``Zorotovic sample'') as well as the confirmed PCEBs from the catalogue of  \citet{2003A&A...404..301R} that were not contained in any of the other published lists (hereafter the ``RK sample''). 

The process for assembling the Gold Sample is illustrated in Table\,\ref{tab:reconcilation}. We first match each system to the closest \textit{Gaia} source within 2\,arcsec and then removed those with distances (\texttt{r\_est} from \citealt{2018AJ....156...58B}) greater than 300\,pc. We then removed any systems that fail the data quality criteria  (\ref{eq:1}) to (\ref{eq:4}) and finally rejected wrongly classified systems, e.g. those with an A to K class companion (which we incorporate in the sample discussed in the next Secion), or those considered to be unreliable. 

The WD+M Gold Sample includes four systems containing strongly magnetic white dwarfs in which accretion of the wind from the M-dwarf companion results in the emission of cyclotron radiation: MQ\,Dra \citep{2008ApJ...683..967S}, WX\,LMi \citep{1999A&A...343..157R}, SDSS\,J030308.35+005444.1   \citep{2013MNRAS.436..241P} and HS0922+1333 \citep{2000A&A...358L..45R}. These systems are considered to be progenitors of magnetic CVs, and hence called pre-polars, or PREPs .  

Finally, whilst we refer to the set binaries assembled here as the WD+M Gold Sample, we  have also included four systems with brown dwarf companions, as their properties and evolution are very similar to those of white dwarfs with low-mass M-dwarf companions: SDSS\,J013532.97+144555.9 (=\,NLTT\,5306, \citealt{2013MNRAS.429.3492S}), WD\,0137--3457, \citep{2006MNRAS.373L..55B},  SDSS\,J141126.20+200911.1  \citep{2018MNRAS.481.5216C} and GD1400, \citep{2005AJ....130.2237F}. 

As mentioned above, we defined 300\,pc as the limit for the WD+M Gold Sample yielding 114 systems, which are listed in Table~ \ref{tab:app_WD+M}.

\subsubsection{Selection effects}
The distribution of the WD+M binaries in the \textit{Gaia} HR diagram is illustrated in Fig.\,\ref{fig:09}, where they define a horizontal ``bridge'' between the white dwarf cooling sequence and the main sequence. The evolutionary interpretation of this bridge is discussed in more detail in Sect.\,\ref{sec:pcebevo}. We discuss a few individual systems to illustrate the effect of the white dwarf and M-dwarf spectral type on the location in the HR diagram. Feige\,24 is a well-studied PCEB \citep{2008ApJ...675.1518K} that contains a particularly hot ($\simeq56\,000$\,K, \citealt{2000ApJ...544..423V}) white dwarf, consistent with having the smallest absolute magnitude among the WD+M Gold Sample. 

In contrast SDSS\,J100609.18+004417.0 contains a a cold (7819\,K) white dwarf and a late-type (M7), low-mass ($0.12\,\mathrm{M_{\sun}}$) secondary \citep{2011A&A...536A..43N}. Both stellar components are intrinsically faint, and the system has one of the largest absolute magnitudes in the Gold Sample. Despite the low white dwarf temperature, the system is located relatively close to the white dwarf cooling track, a consequence of the feeble luminosity of its M-dwarf companion. Finally SDSS\,J013851.54-001621.6 contains an ultra-cool (3570\,K) white dwarf and an M5 companion \citep{2012MNRAS.426.1950P}. The white dwarf contributes hardly anything to the optical flux of the system, and it is hence located within the low-mass end of the main sequence. These two  systems illustrate that the limit for a truly volumetric sample is $\simeq250$\,pc to ensure faint systems such as SDSS\,J100609.18+004417.0 are fully represented (see Fig.\,\ref{fig:09}, left panel).  

A global selection effect that affects the WD+M sample is that it has been primarily constructed from systems identified by SDSS. Consequently, the restricted sky coverage of SDSS has to be taken into account. Based on the sky distribution of $\simeq5.8$ million optical spectra contained within DR16 \citep{2020ApJS..249....3A}, we estimate that SDSS spectroscopy has been obtained over $\simeq29$~per cent of the sky.

The known population of WD+M binaries will be subject to additional selection effects that will affect both the completeness and distribution in the HR diagram parameter space of the Gold Sample defined here. As these systems are, for the vast majority, selected from optical spectroscopy, a key criterion necessary for their identification is that both stellar components contribute noticeable amounts of flux in the visual wavelength range. That immediately implies that a WD+M in which one of the components strongly dominates the optical spectrum will be difficult if not impossible to identify (see Sect.\,\ref{sec:goldwdafgk}). One example consists of hot white dwarfs with very late type M-dwarf companions. The detection of emission lines from the irradiated companion may mitigate the low overall flux contribution of the  companion, although that introduces another selection effect, i.e. the binary separation.  Very cool white dwarfs with relatively early M-dwarf companions form another example. In addition to these intrinsic limitations in the identification of WD+M binaries, there will be additional biases related to the specific observations that were used in a given study, such as the wavelength range covered by the spectroscopy, and brightness limits. 

The following discussion focuses on the SDSS WD+M sample,  which was identified and analysed in a homogeneous fashion, and which makes up the bulk of the Gold Sample. Specifically, the SDSS WD+M sample has a bright-limit of $g\simeq15$ to avoid saturation of the CCDs and contamination of the spectra of fainter objects on adjacent fibres \citep{2002AJ....124.1810S}. This limit will introduce a bias, that we have mitigated by including the systems from  and \citet{2003A&A...404..301R}, \citet{2019MNRAS.484.5362A}, and \citet{2010A&A...520A..86Z}. More subtle, and hard-to-quantify effects arise from the fact that the SDSS sample was obtained using the 640-fibre SDSS and the 1000-fibre BOSS spectrographs sampling a $\simeq7\deg^2$ field-of-view \citep{2000AJ....120.1579Y, 2013AJ....146...32S}. The allocation of these fibres is subject to a complex selection algorithm, that has evolved over the course of the SDSS, and will result in both incompleteness and magnitude and colour-dependent selection effects. \citet{2010MNRAS.402..620R, 2012MNRAS.423..320R, 2013MNRAS.433.3398R, 2016MNRAS.458.3808R} reported spectral types for the white dwarf and M-dwarf components, as well as effective temperatures for a sub-set of the white dwarfs. We discuss the distributions of those parameters in turn.

\begin{table}
\caption{The number of each WD sub-type (where known) in the WD+M Gold Sample. \label{tab:04} }
\begin{tabular}{|l|l|l|}
\hline
Sub-type & &n   \\ \hline
DA       & Only hydrogen features & 94 \\ 
DC       & Featureless continuum & 5  \\ 
WD       & No WD spectral type & 8  \\ 
DB       & Only helium features & 1      \\ \hline
\end{tabular}

\end{table}

\begin{figure*} 
\includegraphics[width=\columnwidth]{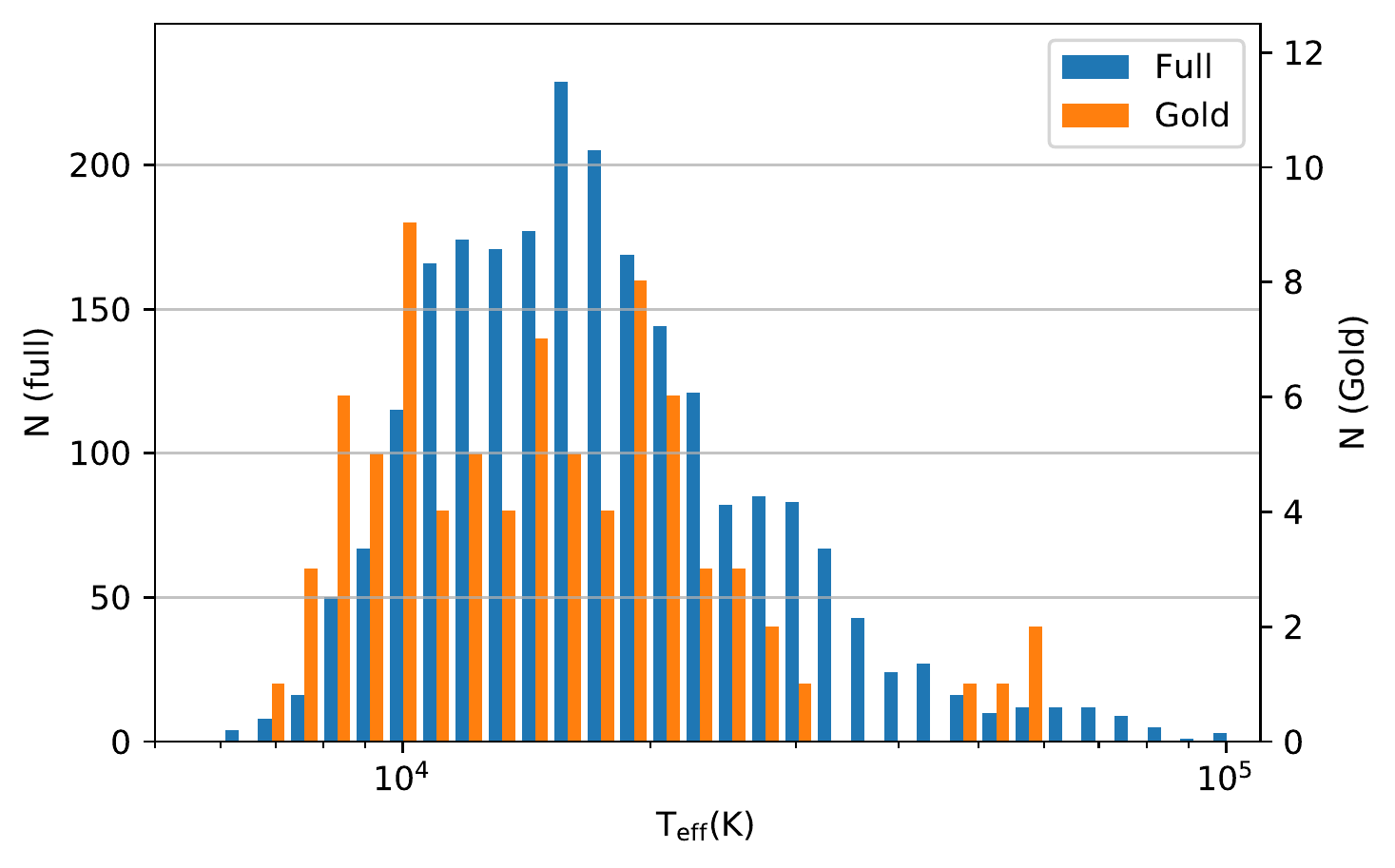}
\includegraphics[width=\columnwidth]{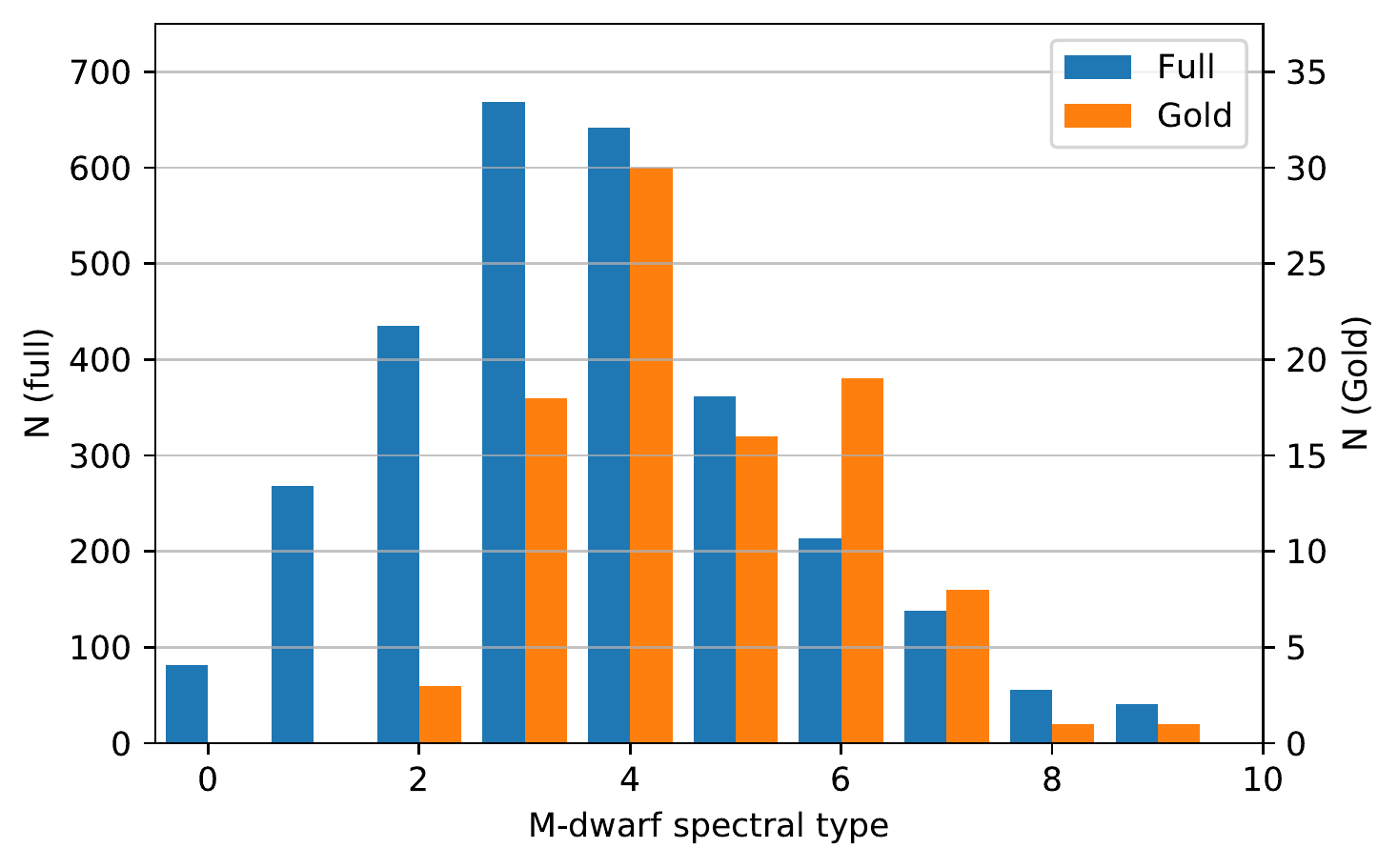}
\caption [.] {\label{fig:11} Left panel: The distribution of the effective temperatures (where known) of the white dwarfs in the WD+M Gold Sample contrasted with the full (not volume-limited) set from from \citet{2016MNRAS.458.3808R}.  Right panel:  The distribution of the spectral types (where known) of the white dwarf companions. The WD+M Gold Sample is contrasted with the full (not volume-limited) set from \citet{2016MNRAS.458.3808R} and shows that the WD+M Gold Sample has a higher proportion of late companions. }
\end{figure*}

\textit{White dwarf spectral type.} The white dwarfs within the SDSS WD+M sample were classified, according to the key features in their optical spectra, into DAs (clearly detected Balmer lines), DB (clearly detected helium lines), DC (featureless continua), and WD (definite blue excess over the M-dwarf, but insufficient to assess the intrinsic spectrum of the white dwarf); the respective numbers of these types are listed in Table\,\ref{tab:04}. The vast majority of the WD+M binaries contain DA white dwarfs, and that fraction may be somewhat higher than among single field white dwarfs in a magnitude-limited sample (e.g. \citealt{2013ApJS..204....5K}). Accretion of the wind of the companion \citep{2006ApJ...652..636D} would relatively rapidly build up a sufficient amount of hydrogen to convert an initially helium-dominated white dwarf into a DA. Figure\,\ref{fig:09} (centre panel) illustrates the distribution of white dwarf spectral types within the \textit{Gaia} HR diagram. As expected, the WD+M systems with barely recognisable contributions from the white dwarf are mostly localised close to or in the main sequence. 

\textit{White dwarf temperature.} The temperature distribution of the white dwarfs within the Gold WD+M sample in the HR diagram is shown in Fig.\,\ref{fig:09} (right panel). Again, as expected, the WD+M binaries with the lowest temperatures have the largest absolute magnitudes. Comparing the white dwarf temperatures in the Gold WD+M sample ($d\la300$\,pc) with those in the full sample (Fig.\,\ref{fig:11}) reveals a bias against hot white dwarfs within the Gold Sample. This bias is a consequence of the brightness limit of the SDSS spectroscopy~--~whilst the actual value of that limit varied throughout the different instalments of the survey, and its diverse sub-programs, very few stars brighter than $g\simeq16$ were observed with the SDSS or BOSS spectrographs, and the hottest white dwarfs will violate that limit at $d<300$\,pc.

\textit{Spectral type of the companion.} \citet{2010MNRAS.402..620R} provide the spectral types of the M-dwarf companions of their sample. Figure\,\ref{fig:11} shows that the distribution of the WD+M Gold Sample is biased in favour of later stars. This again shows the difference between a volume-limited and a magnitude-limited sample, with the intrinsically fainter late spectral types being under-represented in the latter one. A more subtle selection effect is the correlation between donor spectral type and white dwarf temperature: for the WD+M to be detected as such, the white dwarf will need to be brighter, and hence hotter, for earlier type donors. It follows that a magnitude-limited sample will favour intrinsically bright white dwarfs and hence earlier type  donors.

\subsubsection{The Gold Sample as a fraction of the total population}\label{subsubsec:space_density}

The space density of the WD+M Gold Sample is reported in Table~\ref{tab:34} for the three adopted values of the scale height. Remembering the selection effect caused by the SDSS sky coverage these space densities should be multiplied by $\sim$3.4 before making comparisons with the intrinsic space density of WD+M binaries. Few published studies on observational estimates of the space density of WD+M binaries exist, \citet{2003A&A...406..305S}\footnote{The sample of \citet{2003A&A...406..305S} includes a small number of PCEBs with companion spectral types earlier than M.} quote $6-30\times10^{-6}\,\mathrm{pc^{-3}}$ which is substantially higher than the values we derive here~--~however, these authors accounted for the selection effects introduced by the dimming of the white dwarf throughout the PCEB evolution. The WD+M Gold Sample clearly forms a small subset of the underlying population, and is biased in favour of systems with hot white dwarfs and early-type M-dwarfs.

\begin{table}
\caption{The estimated effective volume ($V_\mathrm{eff}$) and space density of the Gold Samples as a function of scale height ($h$). These ignore selection effects such as SDSS sky coverage (see text for details). \label{tab:34}}
\begin{tabular}{|l|l|r|l|l|l|} 
\hline
h (pc) & \begin{tabular}[c]{@{}l@{}}Veff\\ $(10^{6}\,\mathrm{pc^{3}})$ \end{tabular} & \multicolumn{4}{c|}{\begin{tabular}[c]{@{}c@{}}$\rho_0$\\ $(10^{-6}\,\mathrm{pc^{-3}})$ \end{tabular}}  \\ 
\cline{3-6}
       &                                                                   & \multicolumn{1}{l|}{WD+M} & WD+AFGK & CV   & DWD                                              \\ \hline
100    & 46.5  & 2.47  & 0.37    & 3.27 & 1.35                     \\ 
280    & 65.4  & 1.76  & 0.26    & 2.23 & 1.04                     \\ 
500    & 93.3  & 1.23  & 0.18    & 1.63 & 0.73                     \\
\hline
\end{tabular}
\end{table}

\subsubsection{Candidates with no parallax}
\begin{figure*} 
\centering
\begin{minipage}{8.5cm}
\includegraphics[width=8.5cm]{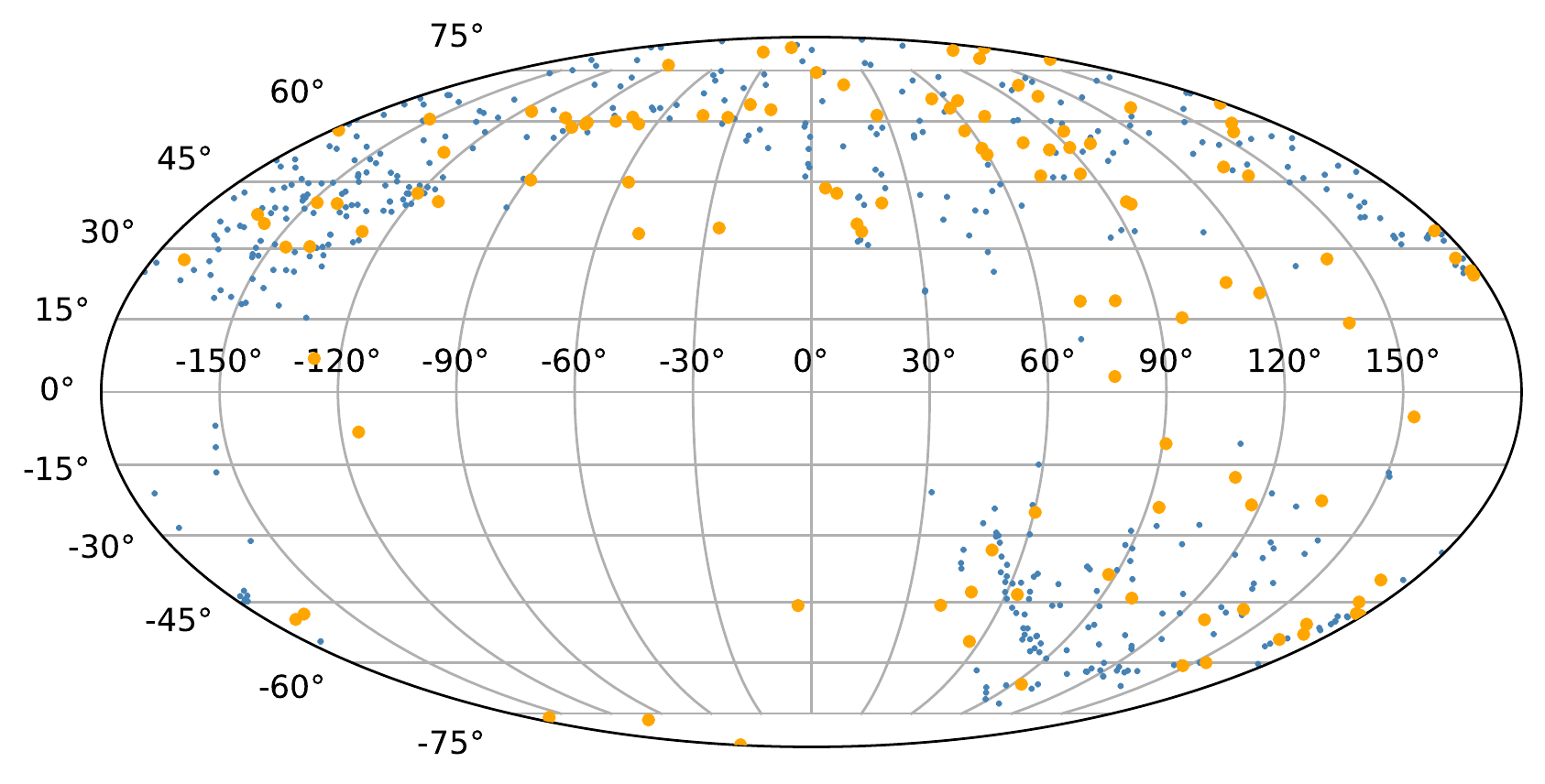}
\end{minipage}%
\begin{minipage}{8.5cm}
\includegraphics[width=8.5cm]{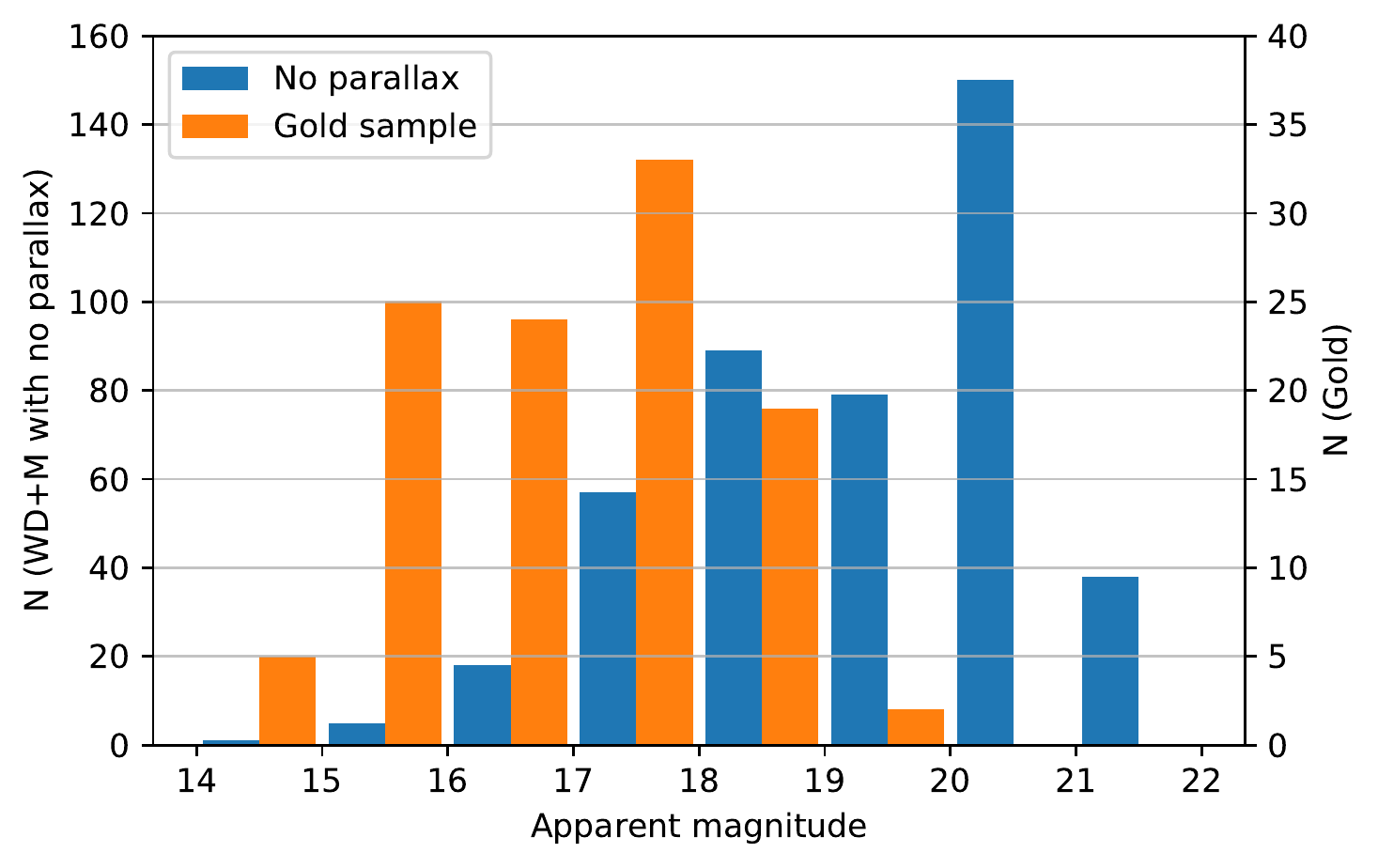}
\end{minipage}%
\caption [.] {\label{fig:18}  Left panel: The distribution of the WD+M Gold Sample (orange) and candidates without a \textit{Gaia} parallax (blue) in Galactic coordinates. Right panel: The $G$-band magnitude distribution of the WD+M Gold Sample compared with the WD+M systems without a parallax. It is evident that most of the systems without a parallax are fainter than the Gold Sample~--~although the parallax incompleteness is already rising when $G\ga17$, which is several magnitudes brighter than \textit{Gaia}'s magnitude limit.
}
\end{figure*}

Of the 3520 candidates (Table\,\ref{tab:reconcilation}) 432 did not have a \textit{Gaia} parallax. We consider the distribution of these 429 systems across the sky area and magnitude range (Fig.\,\ref{fig:18}). Whereas the sky coverage of the SDSS footprint is evident (see \citealt{2017AJ....154...28B}), there is no apparent spatial pattern among the systems without parallax. However, the magnitude distribution reveals a clear drop in the fraction of WD+M candidates with parallaxes towards fainter systems. The parallax incompleteness is already very noticeable for $G\ga17$, which is substantially above the \textit{Gaia} magnitude limit. The reason for the apparent difficulty of \textit{Gaia} in measuring parallaxes of WD+M systems is not clear, but we speculate that some of those systems might be wide binaries that exhibit sufficient astrometric accelerations to foil the DR2 five-parameter astromeric solution. This hypothesis is supported by the large number of systems found between the white dwarf cooling track and the main sequence that have excessive \texttt{ruwe} values (see fig.\,1 in \citealt{2020MNRAS.496.1922B}).

\subsection{PCEBs - White Dwarf + A/F/G/K stars}\label{sec:goldwdafgk}
PCEBs consisting of a white dwarf and a star of spectral type A, F, G and K (hereafter WD+AFGK) present very different observational issues from those with M-dwarf companions as the optical flux of the companion star is so much greater than that of even hot and young white dwarfs. Nevertheless, the process of establishing systems as close WD+AFGK binaries is the same as for the WD+M ones, i.e. firstly identifying candidate systems, and secondly establishing whether they are wide binaries or PCEBs via the appropriate follow-up observations.

\subsubsection{Establishing the Gold Sample}
WD+AFGK binary candidates have been identified by searching for systems with an ultraviolet excess~--~which in most case signals the presence of the white dwarf.  As PCEBs have short periods their confirmation requires the detection of radial velocity variations related to the orbital motion. Given that the companions in these systems totally dominate the optical flux, their spectra are rich in sharp photospheric features, and they are typically relatively bright compared with WD+M systems. 

The identification of WD+AFGK binary candidates has been pioneered by \citet{2016MNRAS.463.2125P} and \citet{2017MNRAS.472.4193R} who cross-matched  AFGK-type stars spectroscopically confirmed by the RAVE and LAMOST surveys respectively, with the \textit{GALEX} ultraviolet photometry. \citet{2016MNRAS.463.2125P}  obtained \textit{HST} spectroscopy of nine of these potential WD+AFGK binaries, all of which contained white dwarfs (see also \citealt{2020arXiv200902968W}).\footnote{This does leave the possibility that some systems are triples. \citet{2020MNRAS.494..915L} estimated this fraction to be less than 10~per cent.}. Follow-up spectroscopy of TGAS-selected WD+AFGK candidates led to the confirmation of 23 PCEBs (\citet{2020arXiv201002885R},  hereafter the ``Ren sample''). Four additional PCEBs were confirmed by \citet{2015MNRAS.452.1754P} and Hernandez et al. (2020, submitted). We included also the well-studied PCEBs V741\,Tau \citep{1970PASP...82..699N} and IK\,Peg \citep{1993MNRAS.262..277W}, as well as the recently discovered GPX-TF16E-48 \citep{2020MNRAS.493.5208K}. 

We also consider in this sample the EL\,CVn binaries, short-period ($\simeq1-3$\,d) binaries containing A/F-type stars with extremely low-mass (ELM) white dwarf companions. These systems are identified as eclipsing binaries \citep{2014MNRAS.437.1681M, 2018MNRAS.475.2560V}. EL\,CVn stars represent a different evolutionary channel from the WD+M and WD+AFGK binaries discussed so far, forming via dynamically stable mass transfer that initiates once the initially more massive star in a main-sequence binary enters the sub-giant branch \citep{2017MNRAS.467.1874C, 2018ApJ...858...14S}. There is strong evidence that most EL\,CVn systems are inner binaries of hierarchical triples \citep{2020arXiv201003507L}. We include the EL\,CVn stars listed by \citet{2003A&A...404..301R} (hereafter the ``RK sample''). As before systems without a \textit{Gaia} match or parallax were dropped and quality filters (Eq.\,\ref{eq:300pc} to \ref{eq:4}) were applied (see Table~\ref{tab:reconcilation})\footnote{In the case of WD+AFGK binaries, the main sequence star is intrinsically so bright that it will be reliably detected to well over 300\,pc. However for compatibility with the other Gold Samples, we use the same limiting volume.}. The WD+AFGK Gold Sample contains 15 systems (Table\,\ref{tab:app_WD+AFGK}).

\subsubsection{HR Diagram} \label{sec:FGKHR}
\begin{figure*} 
\centering
\begin{minipage}{8.5cm}
\includegraphics[width=8.3cm]{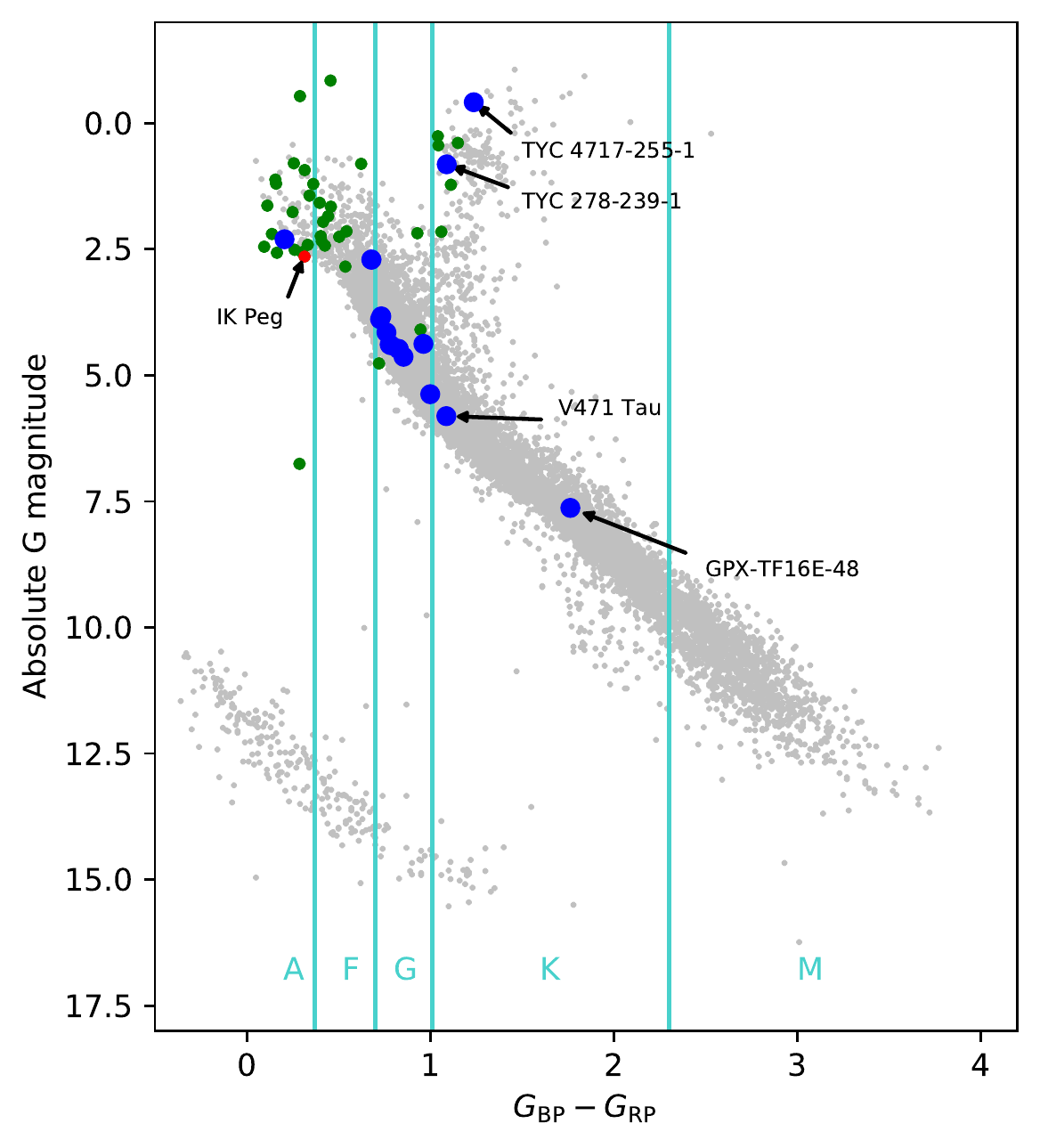}
\end{minipage}%
\begin{minipage}{8.5cm}
\includegraphics[width=8.5cm]{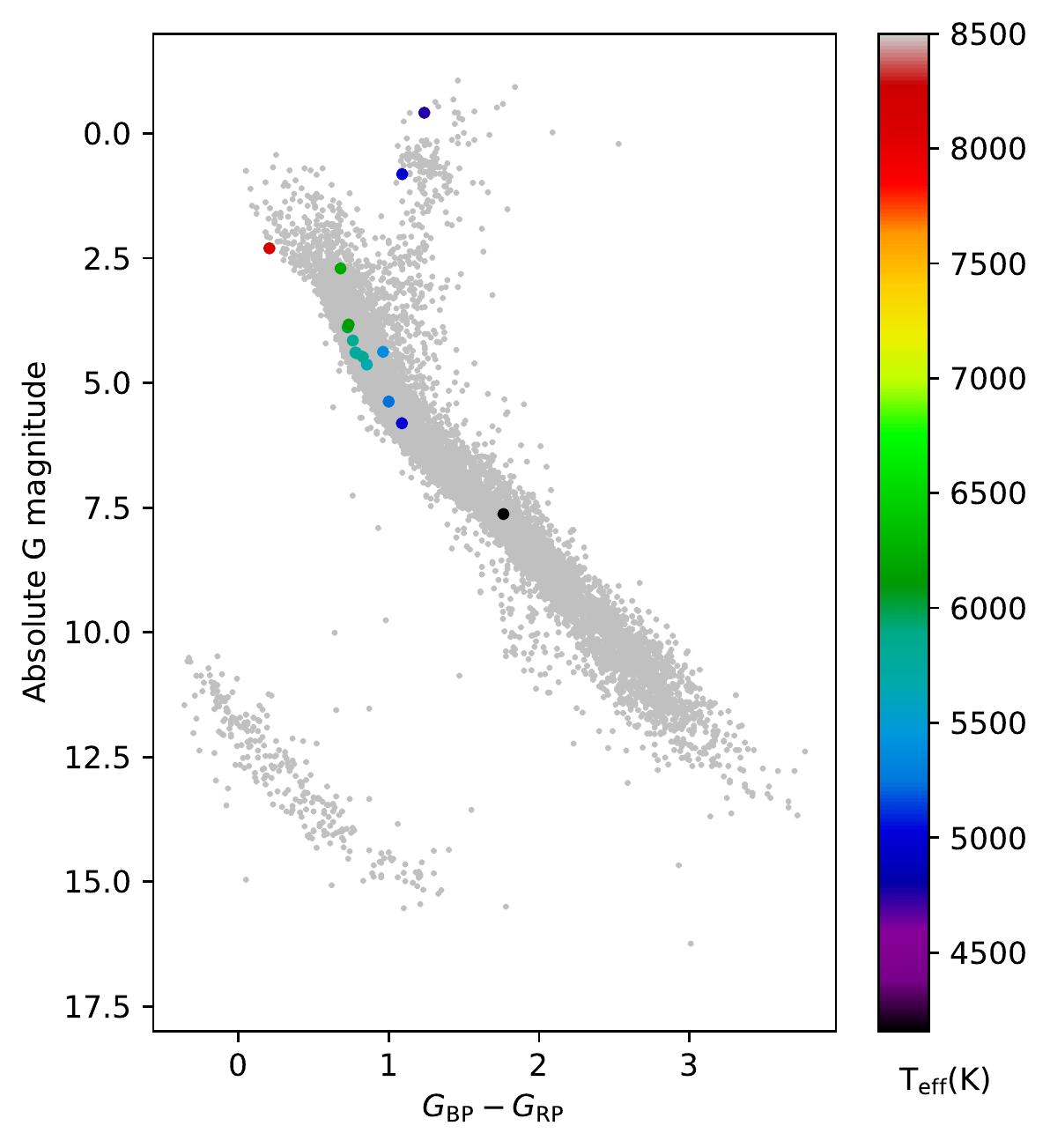}
\end{minipage}%
\caption []{\label{fig:23}  
Left panel: 
The WD+AFGK Gold Sample in the HR diagram contrasting the  WD+AFGK PCEBs over 300\,pc (green), IK\,Peg (red) and  the Gold Sample (blue). The boundaries between spectral types are shown in turquoise. Right panel: The distribution of the effective temperatures of the main-sequence (and sub-giant) companions of the WD+AFGK Gold Sample.
}  
\end{figure*}

The distribution of the WD+AFGK binaries in the \textit{Gaia} HR diagram is shown in Figure~\ref{fig:23}. As the size of this Gold Sample is small (Table\,\ref{tab:reconcilation}), we include in this diagram, purely  for the purpose of visualisation, systems beyond the 300\,pc volume limit; these systems are not included in the Gold Sample. 

WD+A binaries are located at the top of the main sequence, and the intrinsic  brightness of the main sequence companions in these systems results in a significant selection effect: All but one are EL\,CVn binaries with a low mass pre-He white dwarf formed by stable mass transfer rather than in the common envelope phase. These systems have been almost exclusively discovered because of mutual eclipses of the two stellar components. The Gold Sample contains one EL\,CVn star: the prototype itself.

The one non-EL\,CVn system among the WD+A binaries is IK\,Peg. Whereas it is not included in the Gold Sample because of \textit{Gaia} quality issues, it deserves mention as it further underlines the difficulty in finding WD binaries with luminous companions. IK\,Peg had long been known as a single-lined spectroscopic binary with a period of 21.7\,d \citep{1928PDAO....4..171H}, but the white dwarf component was discovered only much later via ultraviolet spectroscopy \citep{1993MNRAS.262..277W}. The fact that IK\,Peg is nearby,  $\simeq47$\,pc, suggests that WD+A PCEBs may be relatively common.

Whereas the Gold Sample contains a number of white dwarf binaries with F- and G-type companions, there is a distinct lack of systems with K-type companions. Noticeable exceptions are V471\,Tau (with a K2-type companion, \citealt{2007AJ....134.1206K}) and GPX-TF16E-48 (with a K7-type companion, \citealt{2020MNRAS.493.5208K}).  This is probably related to selection effects caused by a gap between the two different approaches used to identify white dwarf plus main-sequence binaries (i.e using either optical spectroscopy of ultraviolet excess detections). 

Within the HR diagram (Fig.\,\ref{fig:23}), the majority of the WD+AFGK Gold Sample occupies the expected well-confined area within the main-sequence, albeit two Gold systems are located within the giant branch: TYC~4717-255-1 is a K0~III star \citep{1999MSS...C05....0H} at 299\,pc and TYC~278-239-1 is also a  K0~III star at 159\,pc \citep{2002A&A...386..709F}.

\begin{figure} 
\includegraphics[width=\columnwidth]{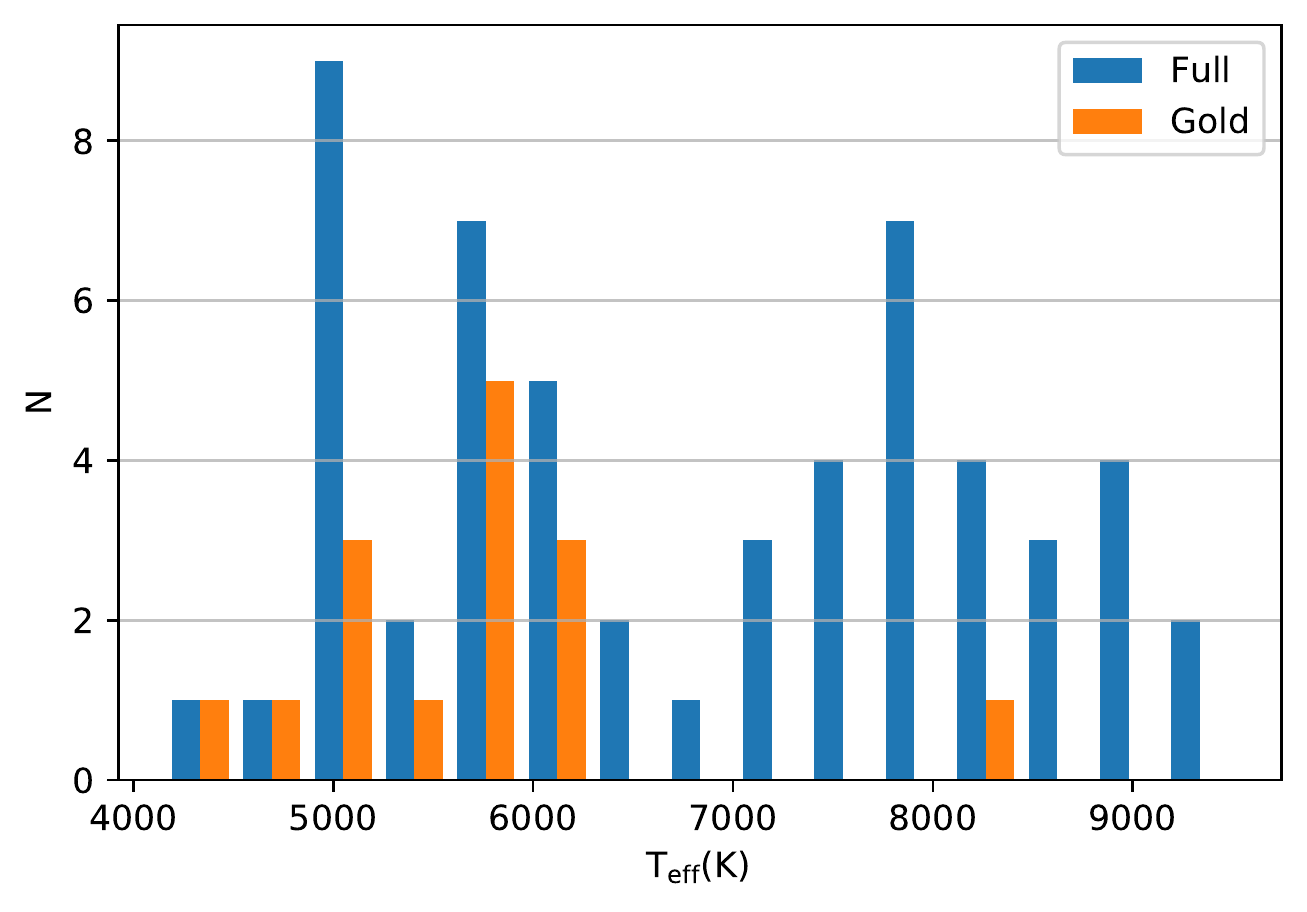}
\caption [.] {\label{fig:24}  The distribution of the effective temperatures of the systems in the WD+AFGK Gold Sample contrasted with the full (not volume-limited) set. The volume-limited sample has a significantly lower average effective temperature.}
\end{figure}

Given that the contribution of the white dwarf to the optical fluxes of the WD+AFGK binaries is negligible, the effective temperatures (taken from from \textit{Gaia} DR2 where available) 

are essentially those of the main sequence or sub-giant stars. The Gold sample is dominated by systems with cooler, G/K-type companions (Fig.\,\ref{fig:24}), with only a single WD+A system, i.e. the prototypical EL\,CVn. This reflects the difficulty in identifying white dwarf binaries with early type companions, the exception being the EL\,CVn systems which are identified photometrically via the detection of eclipses.

\subsubsection{The Gold Sample as a fraction of the total population}
As before (Sect.\,\ref{subsubsec:space_density}) we calculate the space density of this Gold Sample for three different values of $h$ (100, 280, 500\,pc) (see Table\,\ref{tab:34}). These figures compare with $\rho_0\simeq4-10\times10^{-6}\,\mathrm{pc^{-3}}$ for the EL\,CVn stars alone  \citep{2017MNRAS.467.1874C} and a lower limit on WD+FGK binaries of  $\rho_0>1.9\times10^{-6}\,\mathrm{pc^{-3}}$ \citep{2020arXiv201002885R}, both of which suggest that the WD+AFGK Gold Sample forms a small subset of the underlying population. Moreover this subset is heavily biased in favour of systems where the white dwarf signal can be detected, i.e. namely hot white dwarfs, and we conclude that the population of known WD+AFGK binaries is likely the most incomplete out of all the close white dwarf binary sub-types discussed in this paper.

\subsection{Cataclysmic variables}\label{sec:CVs}
Historically CVs, or more specifically CV candidates, were primarily discovered because of their large-amplitude optical variability, either by visual observers, or from the comparison of photographic plates taken at different epochs \citep{2014MNRAS.443.3174B}. Examples of early discoveries include the dwarf nova U\,Gem \citep{1857MNRAS..17..200P} and SS\,Cyg \citep{1896ApJ.....4..369P}. Variability is still a major source of new CV candidate  discoveries, with time-domain surveys such as CRTS \citep{2014MNRAS.441.1186D}, ASAS-SN \citep{2014ApJ...788...48S}, ZTF \citep{2019PASP..131a8002B} and \textit{Gaia} alerts \citep{2013RSPTA.37120239H} continually generating new candidates for investigation. Given that a number of phenomena can be mistaken for CVs, such as minor planets \citep{1996A&A...312..496S} or large-amplitude pulsating stars \citep{1992IBVS.3749....1D}, the unambiguous confirmation of CVs typically requires follow-up spectroscopy \citep{1994A&AS..107..503Z, 2020AJ....159..198S}.

\begin{figure*} 
\centering
\begin{minipage}{8.5cm}
\includegraphics[width=8.5cm]{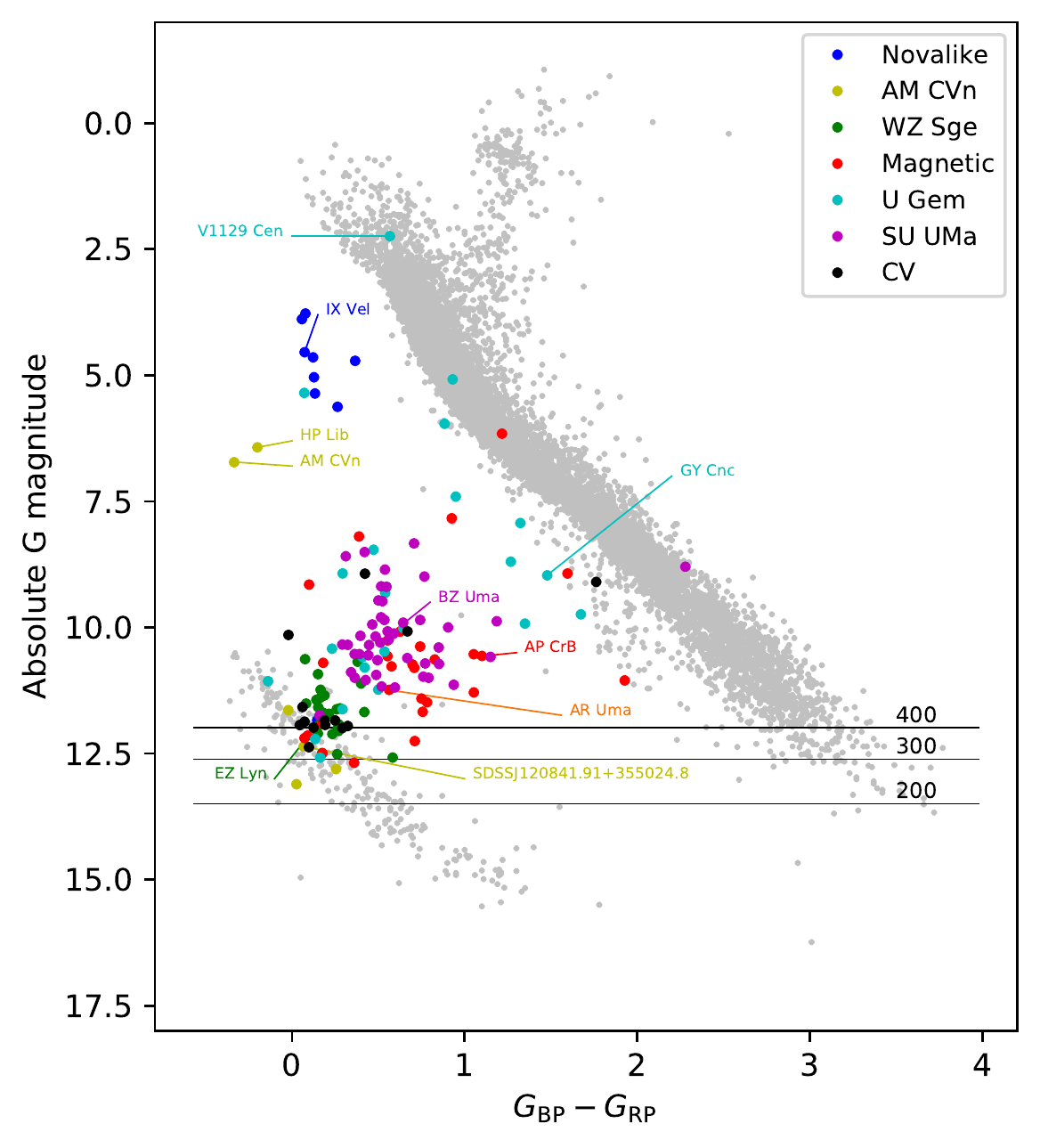}
\end{minipage}%
\begin{minipage}{8.5cm}
\includegraphics[width=8.5cm]{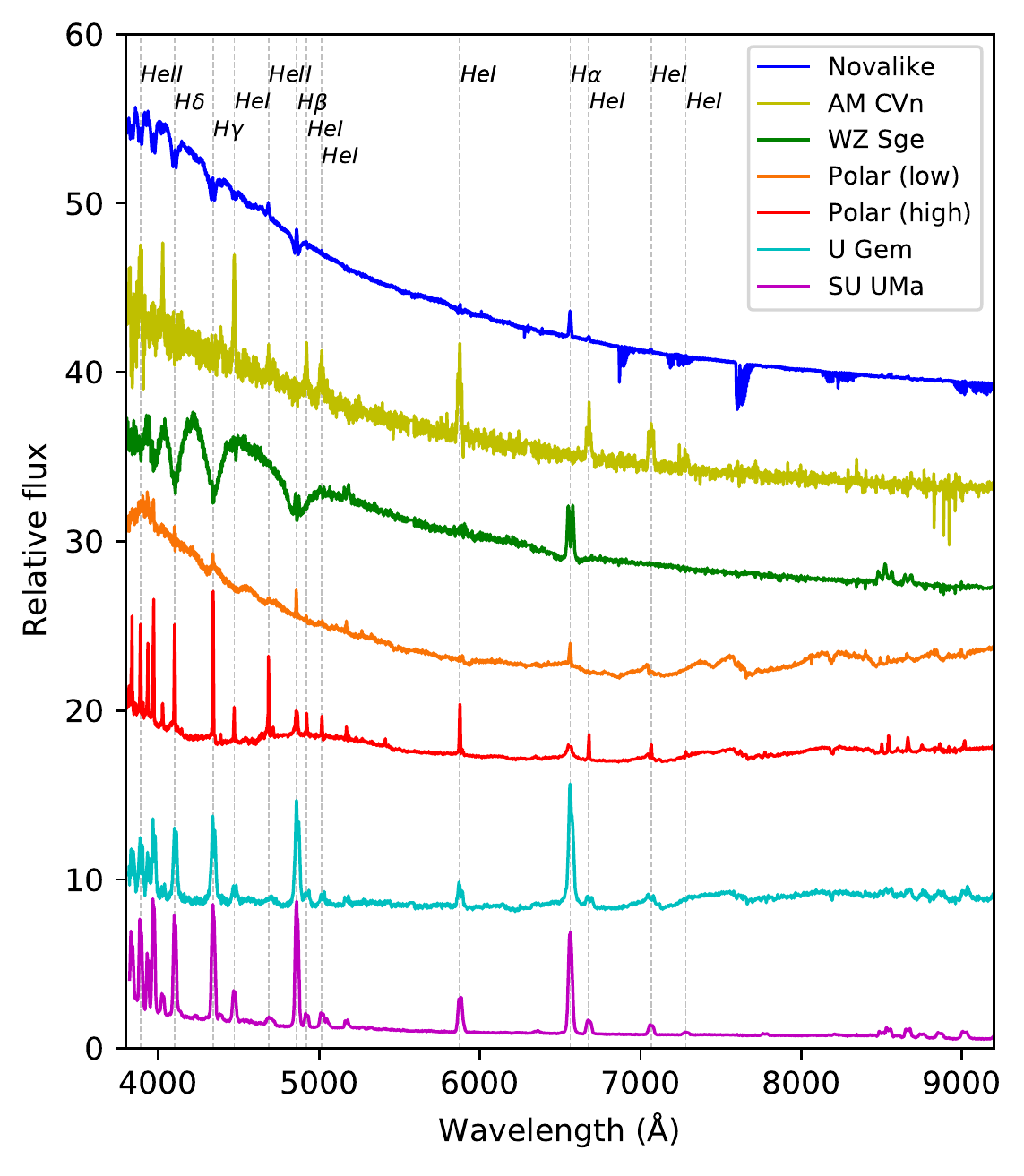}
\end{minipage}%
\caption []{\label{fig:29} Left panel: The HR diagram of the CV Gold Sample shows that the different sub-types cluster in preferred regions of parameter space. There is insufficient data to classify the objects labelled ``CV''. The location of these clusters arises from the different spectral energy distributions of the various sub-classes, which are a function of the properties of both stellar components, and the accretion flow. The minimum absolute magnitude to be detectable at an apparent magnitude of $G=20$ is indicated by grey lines for distances of 200, 300 and 400\,pc. Right panel: Typical examples of the optical spectra for each sub-type. The spectra have been normalised and stacked applying appropriate vertical offsets. The location of each individual CV is identified in the HR diagram. The spectrum of the novalike (=\,IX\,Vel) was obtained using X-Shooter \citep{2011A&A...536A.105V}. The spectra of the AM\,CVn  (=\,SDSS\,J120841.91+355024.8), two polars (AR\,UMa in a low state, and AP\,CrB in a high state), and three dwarf novae of the sub-types U\,Gem, SU\,UMa, and WZ\,Sge (=\,GY\,Cnc, BZ\,UMa, and EZ\,Lyn, respectively) were obtained from SDSS DR16 \citep{2020ApJS..249....3A}. }  
\end{figure*}

CV candidates discovered because of their optical variability are biased towards systems that undergo frequent, and/or large-amplitude brightness changes. Alternative characteristic hallmarks of CVs are their blue colour, Balmer and helium emission, and X-ray emission, which allow their identification in multi-colour photometric surveys \citep[e.g.][]{1986ApJS...61..305G}, spectroscopic surveys \citep[e.g.][]{2006A&A...455..659A, 2009AJ....137.4011S}, and X-ray surveys \citep{2002A&A...396..895S, 2009MNRAS.392..630L}. The relative contributions of the different CV identification methods to the overall population of known CVs, and their affinity with different CV sub-classes are discussed by \citet{2005ApJ...629..451G}. Despite all efforts, CVs that mimic either single white dwarfs, or main-sequence stars, and have low accretion rates remain very difficult to discover (e.g. \citealt{2014ApJ...790...28R, 2019ATel12936....1T, 2019MNRAS.489.1023Y}).

\subsubsection{Establishing the Gold Sample} \label{sec:gs}
An extensive literature review was conducted as part of \citet{2020MNRAS.494.3799P} to  obtain a comprehensive inventory of both confirmed CVs and CV candidates. This list has been updated for new CV candidates that have been discovered since that study, and contains 5192 systems of which 4042 have a \textit{Gaia} parallax (see Table~\ref{tab:reconcilation}). Applying the distance limit we were left with 305 CV candidates. We then applied the quality filters outlined in Sect.\,\ref{sec:gaia} resulting in 201 systems.

As discussed by \citet{2020MNRAS.494.3799P}, a fraction of CV candidates reported in the literature are contaminants, including flaring M-dwarfs, young stellar objects, single white dwarfs and detached binaries being mistakenly identified as CVs. In order to establish the CV Gold Sample, we therefore proceeded to validate the CV nature of these 201 candidates. 

We began by adopting the CV classification for the 42 systems with $d\la150$\,pc, that were already vetted by \citet{2020MNRAS.494.3799P}, as well as 139 with $150\la d\la300$\,pc that were in the catalogue of bona-fide CVs compiled by \citet{2003A&A...404..301R}. We then extracted light curves, where available, for the 124 remaining candidates from ZTF \citep{2019PASP..131a8003M}, ASAS-SN \citep{2019MNRAS.485..961J,2019PASP..131a8003M} and CRTS \citep{2019PASP..131a8003M} together with spectra from SDSS \citep{2013AJ....146...32S} and LAMOST \citep{2012arXiv1206.3569Z}. We found a total of 152 CVs with reliable spectra and a history of variability and these are henceforth referred to as the CV Gold Sample (Table\,\ref{tab:app_CV}). 

\begin{figure*} 
\includegraphics[width=\columnwidth]{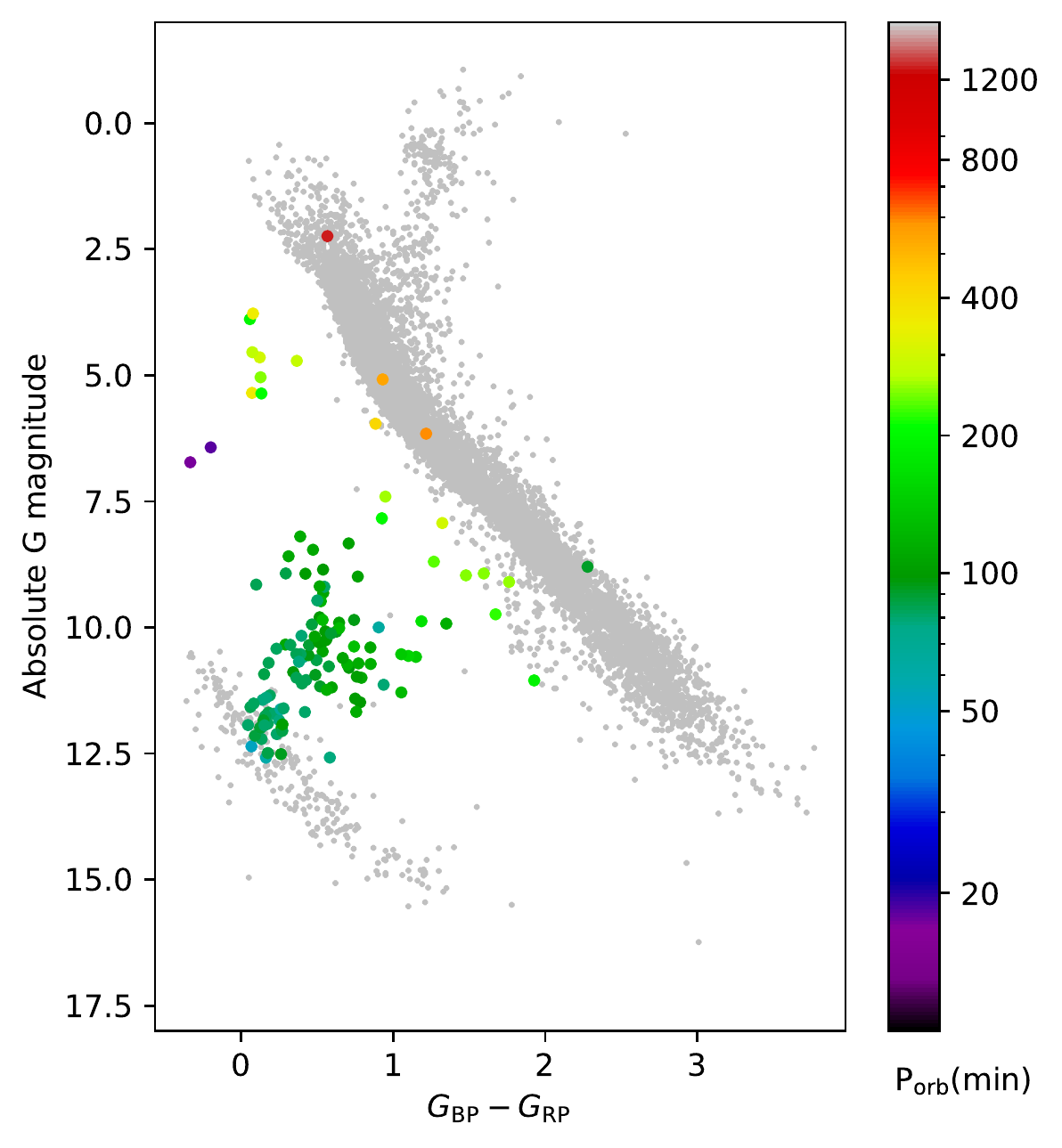}
\includegraphics[width=\columnwidth]{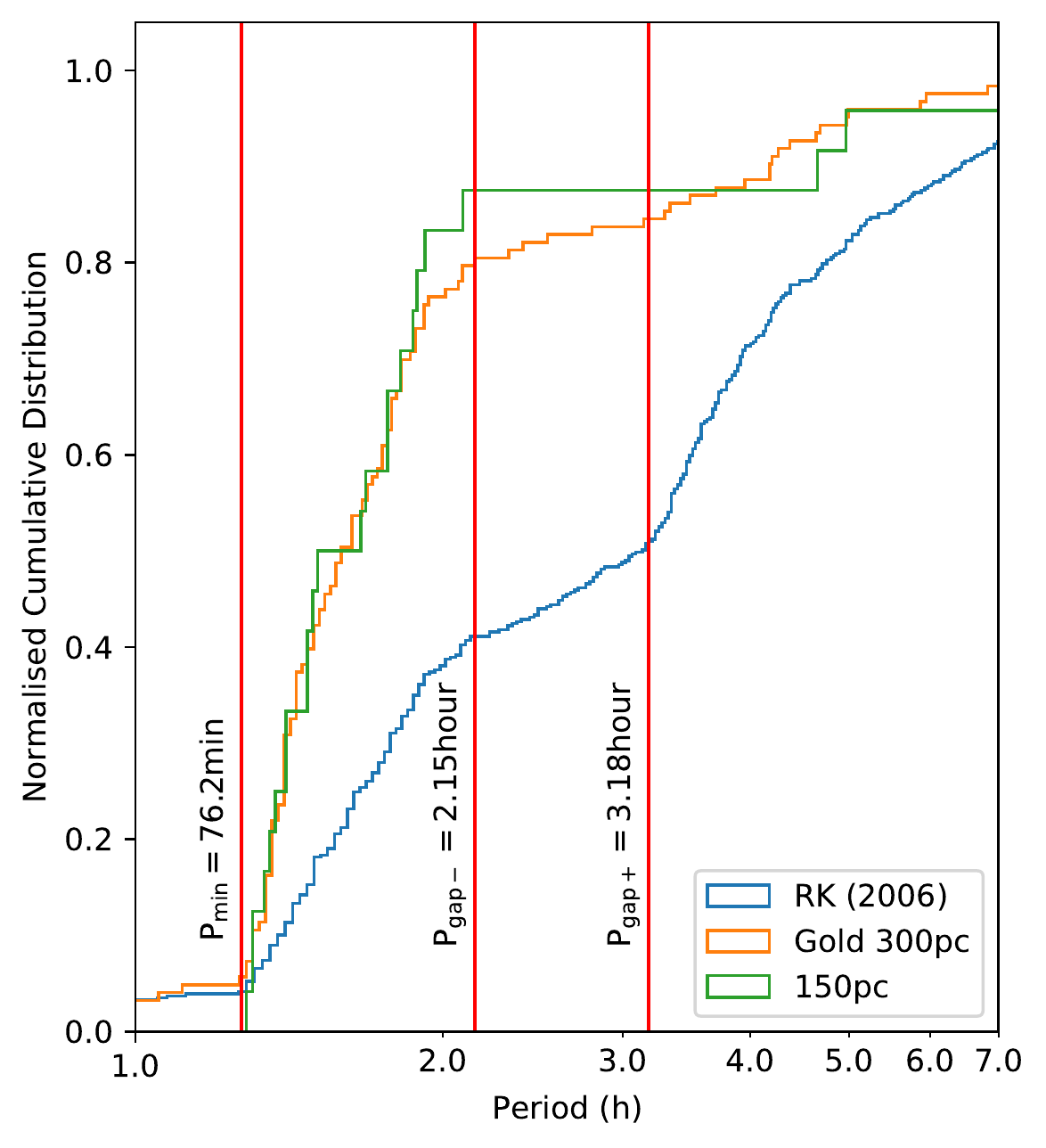}
\caption [.] {\label{fig:16} Left panel: 
An HR diagram showing the distribution of CVs in the Gold Sample with their measured orbital periods colour-coded (in minutes). As CVs evolve towards shorter periods, they move from the main sequence towards the white dwarf cooling track. Right panel: Comparison of cumulative orbital period distributions of the 150\,pc sample (green, \citealt{2020MNRAS.494.3799P}), the 300\,pc CV Gold Sample (orange) and the \citet{2003A&A...404..301R} sample (blue -  we use V7.6 which enables  a direct comparison with fig.\,4 in \citealt{2006MNRAS.373..484K}), which is a heterogeneous mix of CVs identified by all identified selection methods. The period gap starting at  two hours is very evident in each case.}
\end{figure*}

The CVs in the Gold Sample have reliable parallax measurements with an average distance uncertainty (based on the estimated errors in \citealt{2018AJ....156...58B}) of 2.7~per cent. 

In passing we note that we have excluded $\simeq50$ potential CV candidates from the Gold Sample primarily due to a lack of spectroscopy. Follow-up observations are planned and will be reported in a subsequent paper. We also expect that a number of the CV candidates rejected due to poor data quality will be confirmed by \textit{Gaia} EDR3 astrometry. 

\subsubsection{The CV Gold Sample in the \textit{Gaia} HR diagram}
The distribution of the CV Gold Sample within the \textit{Gaia} HR diagram is shown in Fig.\,\ref{fig:29}. We re-iterate that the CV Gold Sample is not intended to be complete, but to establish the parameter space that CVs inhabit, and to assess how their location within this parameter space links to their physical properties (the evolution of CVs across the HR diagram is discussed in more detail in Sect.\,\ref{sec:evolution}). We review how well the CV Gold Sample meets those goals in four ways~--~firstly by checking that the variety of sub-types are well-represented, secondly by checking that there is a spread in evolutionary states (for which orbital period is a proxy), thirdly by comparing the number of CVs in the Gold Sample with the estimated size for a complete 300\,pc CV sample, and lastly by investigating whether the CV candidates from our initial list that do not have a parallax introduce some major bias in the CV Gold Sample.

\paragraph{CV subtypes.} The CV Gold Sample contains examples of all six major different sub-types, their respective numbers are listed in Table~\ref{tab:07}. The different sub-types form clearly distinct clusters in the \textit{Gaia} HR diagram (Fig.\,\ref{fig:29}, left panel)\footnote{A similar HR diagram has been shown by \citet{2020MNRAS.492L..40A}, presenting overall similar features. However, these authors included a much larger number of CVs with much looser selection criteria than those defining the Gold Sample. Consequently, the clustering in their diagram is less pronounced.}, and the location of these clusters closely links to the spectro-photometric properties of each sub-type (Fig.\,\ref{fig:29}, right panel). 

The novalike variables are characterised by stable high mass transfer rates and hot accretion discs that dominate the optical emission of these systems. Correspondingly, their cluster is located at small absolute magnitudes and blue colours. Two AM\,CVn stars, HP\,Lib and the prototype AM\,CVn itself \citep{1972ApJ...175L..79F} are novalike variables with stable, hot helium discs that are somewhat bluer and fainter than their cousins with hydrogen-rich discs. 

The three sub-classes of dwarf novae exhibit clear trends in their location within the HR Diagram. The U\,Gem type dwarf novae have lower mass transfer rates than the novalike variables, their quiescent discs are therefore less luminous, and their donor stars can contribute noticeably to the optical flux of the systems, which is reflected by a number of these systems being located close to the main sequence. The donors and accretion discs in the SU\,UMa systems are less luminous than those in U\,Gem-type dwarf novae, hence they are concentrated closer to the white dwarf cooling sequence. Finally, the WZ\,Sge type dwarf novae have the dimmest accretion discs and donor stars, in fact some of them contain brown dwarf companions (GD552 \citealt{2008MNRAS.388..889U}, QZ\,Lib \citep{2018MNRAS.481.2523P}, 1RXS\,J105010.3--140431 \citep{2001A&A...376..448M}, SDSS J102905.21+485515.2 \citep{ 2016AJ....152..226T}, SDSS\,J143317.78+101123.3 \citep{2016Natur.533..366H}, SSS\,J122221.7--311525 \citep{2017MNRAS.467..597N}). WZ\,Sge are therefore inherently the faintest CVs, and merge into the white dwarf cooling sequence. 

A noticeable outlier among the dwarf novae is V1129\,Cen, a long-period ($\simeq21.4$\,h) CV that exhibits low-amplitude ($\simeq0.6$\,mag) outbursts about once per year, but no emission lines are detected in its spectrum \citep{2017NewA...57...51B}. The donor is a luminous F-type+ star, explaining the location in the HR diagram.  

\begin{figure*} 
\begin{minipage}{8.5cm}
\includegraphics[width=8.5cm]{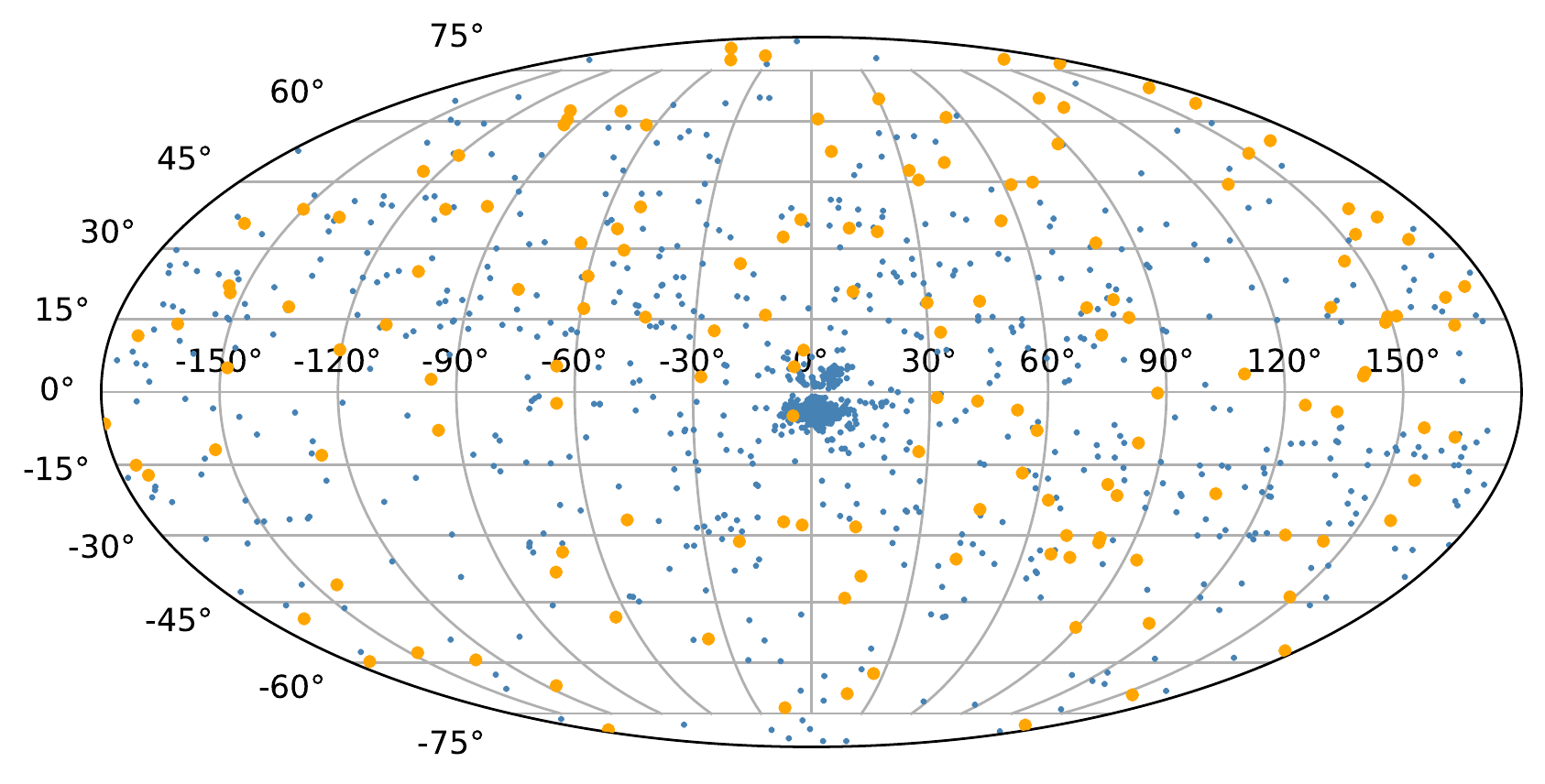}
\end{minipage}
\begin{minipage}{8.5cm}
\includegraphics[width=8.5cm]{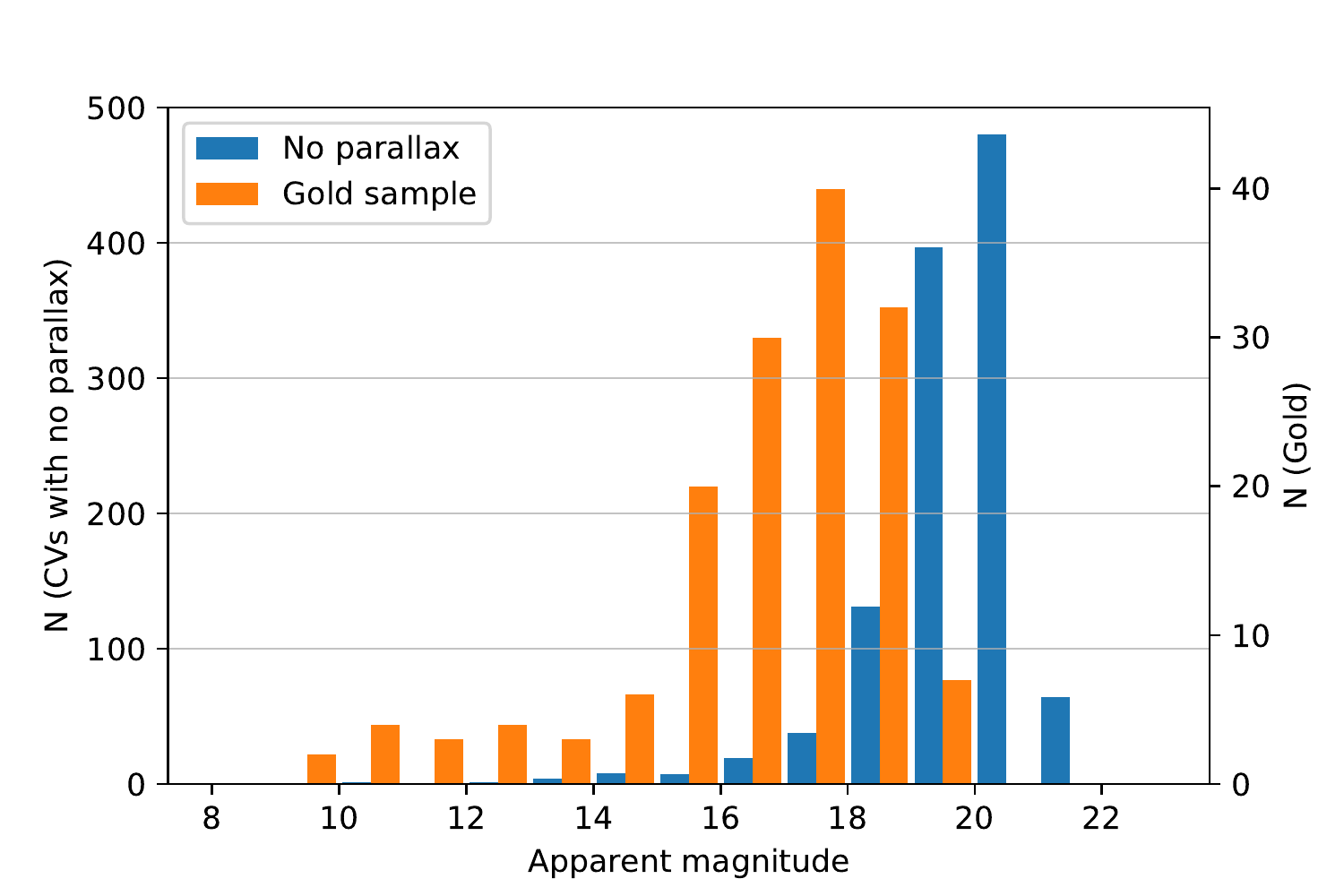}
\end{minipage}
\caption [.] {\label{fig:06}  Left panel: The distribution of the CV Gold Sample (orange) and the CV candidates without a valid parallax (blue) in Galactic coordinates. The concentration of systems with no parallax close to the Galactic centre is very noticeable,  with $\simeq 44$~per cent contained within the area bounded by $-12 < \tt{l} < 12$ and $-8 < \tt{b} < 5$. Most ($\simeq 79$ per cent) of this concentration were identified by the OGLE survey \citep{2015AcA....65..313M}.  Right panel: The  distribution of apparent magnitudes of the Gold Sample contrasted with the CVs without a parallax. It is evident that most of the CVs without a parallax have apparent magnitudes $\ga19$, i.e. near the limiting magnitude of \textit{Gaia}.}
\end{figure*}

Magnetic CVs have a larger spread due to their more varied spectral appearances. During states of low mass transfer they resemble the detached WD+M binaries whilst during high states their optical spectra can contain cyclotron emission lines \citep{1988prco.book..199W} that will affect their colour and absolute magnitude.

Overall, the CV Gold Sample contains a representative mix of the different CV sub-types that well sample the spread of the underlying CV population in the HR diagram. 

\begin{table}
\caption{The number of each type in the CV Gold Sample together with the range of their orbital periods. }
\label{tab:07}
\begin{tabular}{|l|l|l|l|l|}
\hline
\multirow{2}{*}{Type} & \multirow{2}{*}{N} & \multicolumn{3}{c|}{Period (hours)} \\ \cline{3-5} 
                      &                    & Median       & Min       & Max        \\ \hline

Novalike  &  9  &  4.42  &  1.51  &  5.88  \\
AM\,CVn  &  6  &  0.31  &  0.29  &  0.88  \\
WZ\,Sge  &  26  &  1.37  &  1.12  &  1.82  \\
Magnetic  &  28  &  1.89  &  1.37  &  9.88  \\
U\,Gem  &  23  &  2.09  &  0.99  &  21.43  \\
SU\,UMa  &  45  &  1.63  &  1.07  &  2.81  \\
CV  &  14  &  1.44  &  1.3  &  4.26  \\ \hline
\end{tabular}
\end{table}

\paragraph{Distribution of periods.}
As outlined earlier, the evolution of CVs is driven by orbital angular momentum loss, which results in their periods decreasing down to the period minimum before reversing to longer periods as the donor stars become degenerate. The orbital periods of the Gold Sample  (Fig.\,\ref{fig:16}) span the typical range occupied by CVs, $\simeq80$\,min to $\simeq600$\,min, and the distribution of the periods broadly resembles that of the volume-limited 150\,pc sample \citep{2020MNRAS.494.3799P}. However, both the distributions of the 150\,pc and 300\,pc Gold Sample differ markedly from that of the overall population of known CVs, drawn from the \citet{2003A&A...404..301R} catalogue, which displays a steep increase in the number of CVs with periods in the range $3-4$\,h. This difference highlights the importance of using a volume-limited sample for population studies, as intrinsically bright novalikes (with mass transfer rates that in fact exceed the predictions of the standard CV evolution model - see  \citealt{2009ApJ...693.1007T, 2011ApJS..194...28K}) are detected out to much larger distances than other CVs and dominate the period distribution of the \citeauthor{2003A&A...404..301R} sample. 
 
We note in passing that the upper edge of the canonical period gap at $\simeq3$\,h is much less pronounced in the  volume-limited sample when compared to the overall population of CVs in the RK catalogue, which is subject to a multitude of selection effects. This underlines once more that great care has to be taken when using heterogeneous CV samples as the underpinning of generic CV evolution models. 

\paragraph{The Gold Sample as a fraction of the total population.}
In the same manner as the other Gold Samples, we compute the space density of the Gold CVs for three different values of $h$ (100, 280, 500\,pc), resulting in $\rho_0\simeq1.6-3.3\times10^{-6}\,\mathrm{pc^{-3}}$ (see Table~\ref{tab:34}). These values should be compared with $\rho_0=4.8\times10^{-6}\,\mathrm{pc^{-3}}$ for the volume-limited 150\,pc sample \citep{2020MNRAS.494.3799P}, and suggest that the CV Gold Sample represents  a substantial fraction ($\simeq$\,50~per cent) of the underlying population within 300\,pc.

\subsubsection{Candidates with no parallax}
Among the 5192 CV candidates we began with, 1150 have no \textit{Gaia} parallax, and an important question is whether any of those are within 300\,pc. A characteristic of \textit{Gaia} is that the $G$-band magnitude of variable systems with only a 2-parameter solution (i.e. no parallax) will tend to be the (brighter) outburst magnitude not the (fainter) quiescent one. Figure 10.10 in \citet{2018gdr2.reptE..10A} illustrates this effect, and upon closer inspection of a sample, we have found that the majority of CVs with $G$-band magnitudes in the range $\simeq17-20$ (Fig.\,\ref{fig:06}, left panel), and no parallax have indeed fainter quiescent magnitudes,  $\ga20$, in SDSS and PanSTARRS1 imaging, suggesting that the \textit{Gaia} detections were obtained during outburst(s) or high states.

For an assumed distance $\la300$\,pc, these very faint quiescent apparent magnitudes would imply that these systems are intrinsically extremely faint with low mass transfer rates, cool white dwarfs and very low-mass donors. Whereas we cannot rule out that a small number of intrinsically faint WZ\,Sge type dwarf novae with distances near the volume-limit fall into this category, inspection of the spatial distribution of the 1150 CVs without parallaxes shows a strong concentration towards the Galactic centre (Fig.\,\ref{fig:06}, right panel). CVs within a 300\,pc volume limit are expected to be approximately isotropic in their sky distribution, and we conclude that the majority of CVs without a parallax are beyond our adopted distance limit.

\subsection{Double white dwarfs}\label{sec:ddgold}
Over recent years many DWDs have been documented in the literature. Volumetric surveys of white dwarfs encompassing 20\,pc \citep{2018MNRAS.480.3942H} and 40\,pc \citep{ 2020MNRAS.499.1890M}  contain a number of confirmed DWDs and these were supplemented by a collection of systems from various other publications, which include most of the known DWDs. As before (see table \ref{tab:reconcilation}) we remove duplicates and systems for which there is no \textit{Gaia} parallax or which fail our quality criteria outlined in Sect.\,\ref{sec:gaia}. The remaining 67 form the Gold Sample (Table\,\ref{tab:App_DWD}).

\subsubsection{\label{sec:dwd_hrd} The DWD Gold Sample in the \textit{Gaia} HRD}
\begin{figure} 
\includegraphics[width=\columnwidth]{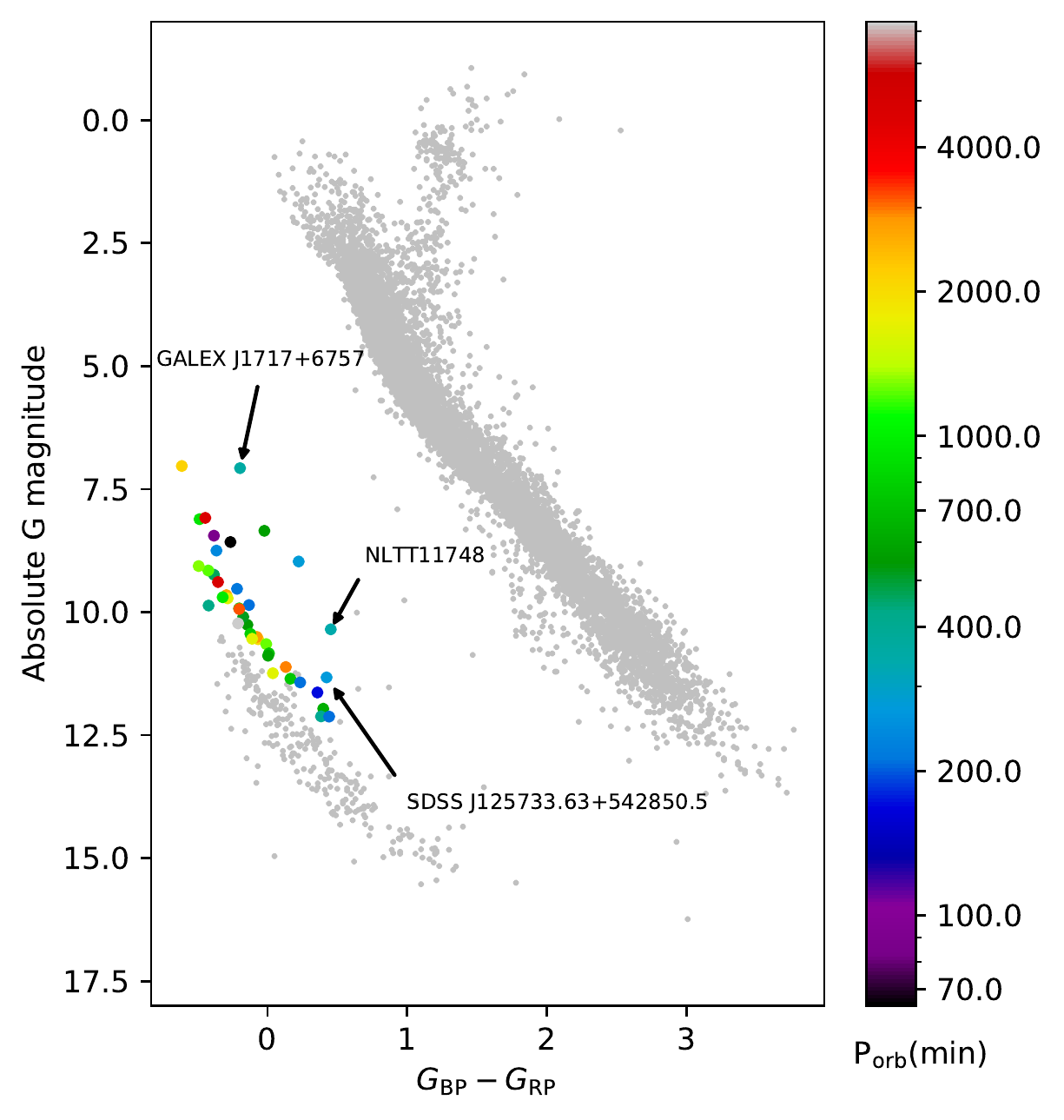}
\caption [.] {\label{fig:33} 
The Gold Sample of DWDs with known orbital periods (colour-coded) are located slightly above the general white dwarf cooling sequence in the HR diagram, reflecting the combined luminosity of the two white dwarfs.
The lack of DWDs at the faint end of the white dwarf sequence is an observational selection effect.}
\end{figure}

A system consisting of two white dwarfs typically has a similar colour to a single white dwarf but is somewhat (up to $\simeq0.75$ magnitudes depending upon their temperatures and to a lesser extent masses) brighter and therefore appears slightly higher in the HR diagram (Fig.\,\ref{fig:33}).  The four outliers bounded by NLTT11748  \citep{2014ApJ...780..167K,2010A&A...516L...7K} and GALEX\,J1717+6757 \citep{2011ApJ...737L..16V} include an extremely low mass (ELM) white dwarf \citep{2020ApJ...889...49B,2020ApJ...894...53K}. ELM white dwarfs (typically $\sim$ 0.2\,M$_{\sun}$) have greater luminosities than more massive white dwarfs because of their larger radii (\begin{math}R\propto  M^{-1/3} \end{math}). This increased luminosity therefore causes them to appear above the white dwarf cooling sequence in the HR diagram. 

The lack of DWDs at the cool end ($T_\mathrm{eff}\la7000$\,K) of the white dwarf sequence is a selection effect. The vast majority of DWDs were identified as spectroscopic binaries \citep[e.g.][]{2005MNRAS.359..648M, 2020A&A...638A.131N}, only a very small were photometrically detected because of eclipses \citep[e.g.][]{2010ApJ...716L.146S} or because of inconsistencies in their spectroscopic features \citep{1990ApJ...361..190B}. At low temperatures, the Balmer lines weaken end eventually disappear, therefore the confirmation of DWD candidates (e.g. selected because they are over-luminous, e.g. \citealt{2012ApJS..199...29G, 2018MNRAS.480.3942H}) becomes very difficult for cool white dwarfs. 

Only 38 of the DWDs in the Gold Sample have an orbital period measurement, and there is no apparent correlation between period and position in the HR diagram (Fig.\,\ref{fig:33}).

\begin{figure} 
\includegraphics[width=\columnwidth]{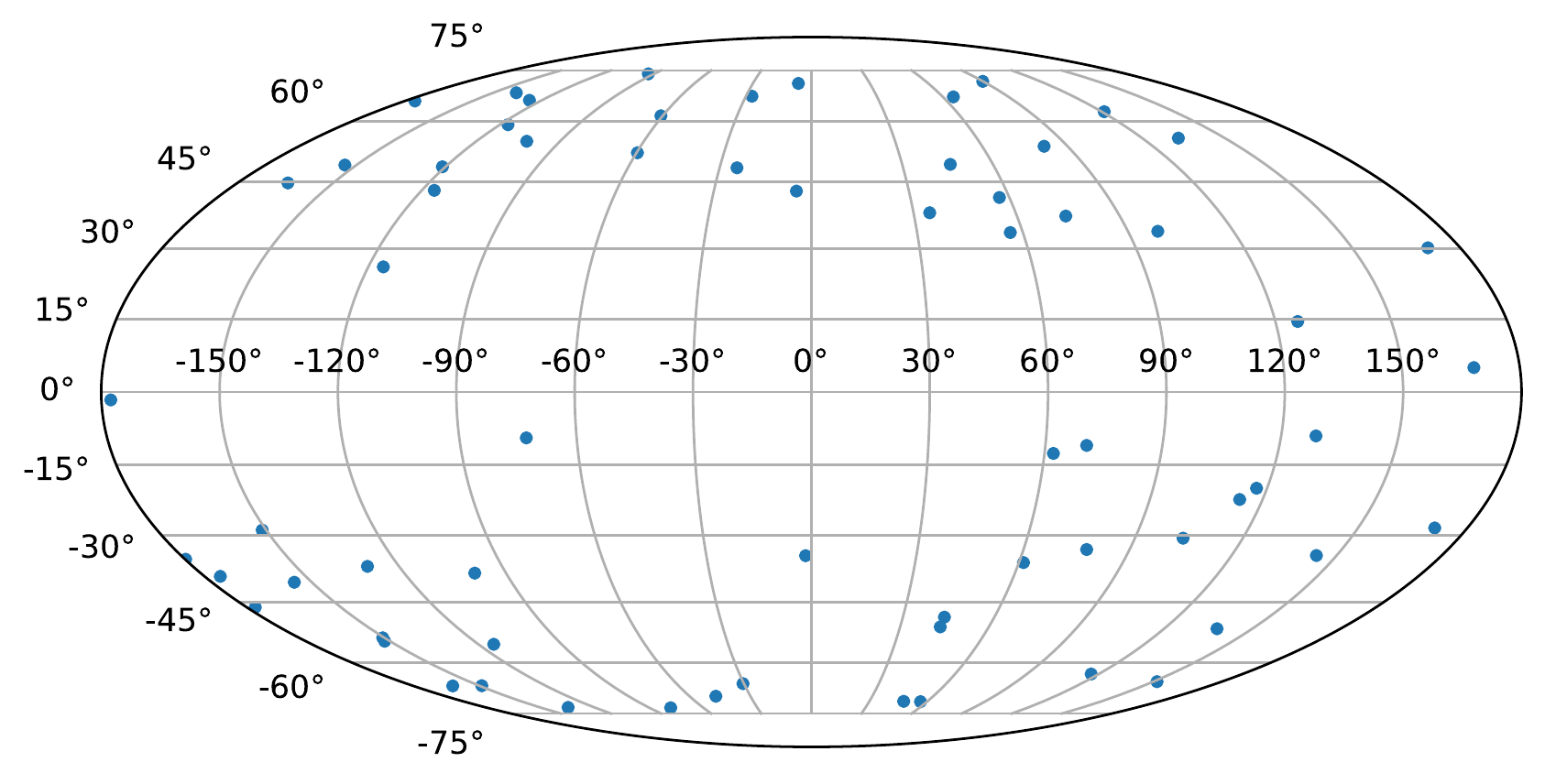}
\caption [.] {\label{fig:DWD sky map} 
The distribution of the DWD Gold Sample in Galactic coordinates. They appear to be evenly distributed away from the Galactic plane. The lack of DWDs near the Galactic plane is a selection effect due to the difficulty in identifying white dwarf candidates in  crowded regions.}
\end{figure}
\subsubsection{The Gold Sample as a fraction of the total population}
As before (Sect.\,\ref{subsubsec:space_density}) we calculate the space density for three different values of h (100, 280, 500\,pc) (see Table\,\ref{tab:34}). These figures compare with $>0.62\times10^{-3}\,\mathrm{pc^{-3}}$ from \citet{2016MNRAS.462.2295H} which demonstrates that the DWD Gold Sample forms a tiny subset of the underlying population. This is partly a reflection of the inherently low luminosity of these systems~--~\citet{2016MNRAS.462.2295H} derived their estimates from a volume-limited survey of 25\,pc~--~and partly the difficulty in distinguishing DWDs from single white dwarfs.

\subsubsection{Candidates with no parallax}
Only six of the 167 candidates had no \textit{Gaia} parallax and in each case their $G$-band magnitude was larger than any of the members of the Gold Sample. It is therefore likely that they are all at a distance of more than 300\,pc. 

This volume-limited sample is weighted towards normal mass white dwarfs rather than the rare, but more luminous ELM DWDs that are detectable out to larger distances, again highlighting the difference between a volume-limited survey and a magnitude-limited one.

\section{Evolution of white dwarf binaries within the HR diagram.}\label{sec:evolution}
The discussion so far has focused on establishing and validating the various white dwarf binary sub-samples, and to discuss their observational properties. Here we will briefly investigate the evolution of white dwarf binaries across the HR diagram, as well as the effects of systems moving from one sub-sample into another one. 

\begin{figure*} 
\includegraphics[width=\columnwidth]{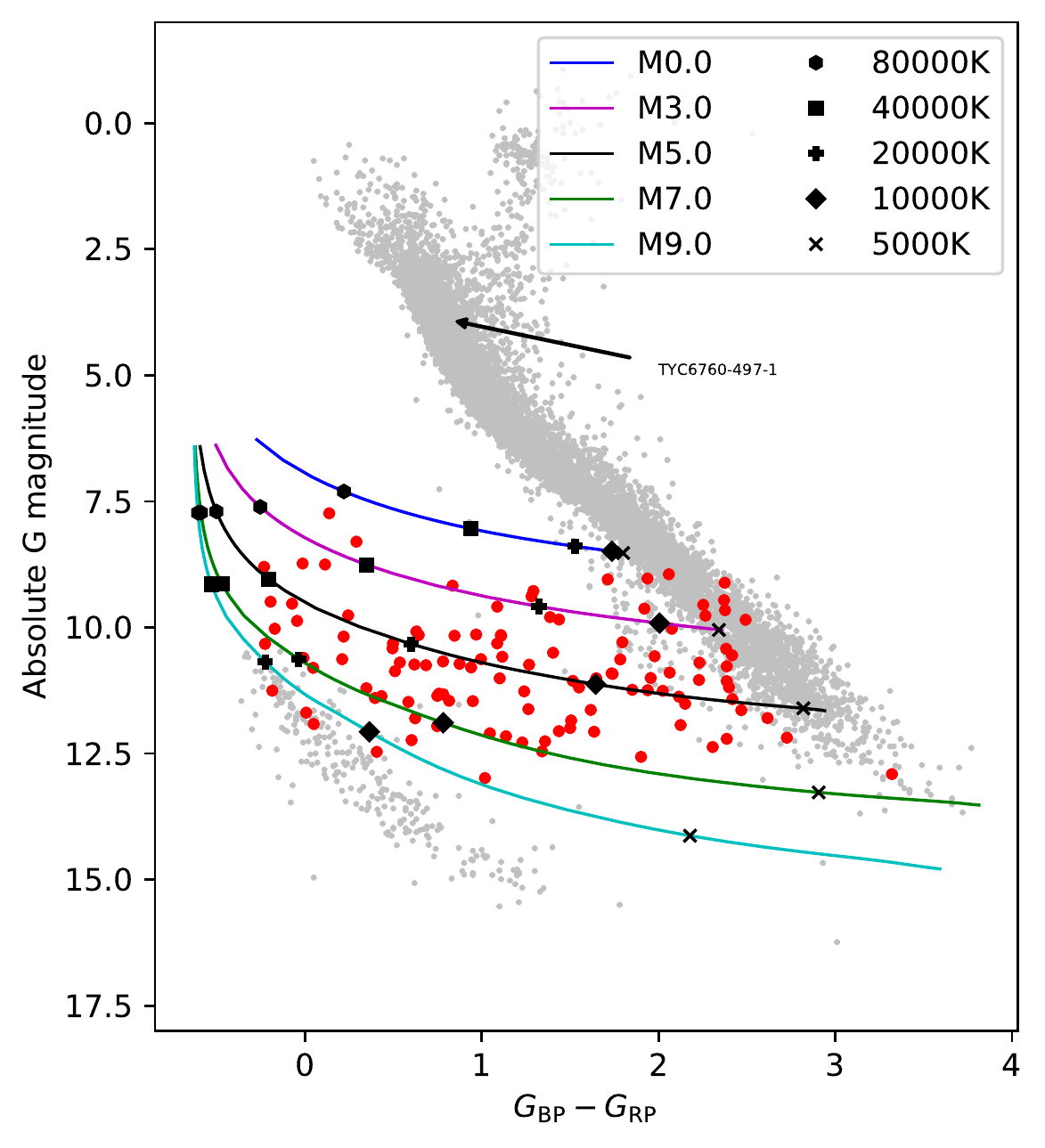}
\includegraphics[width=\columnwidth]{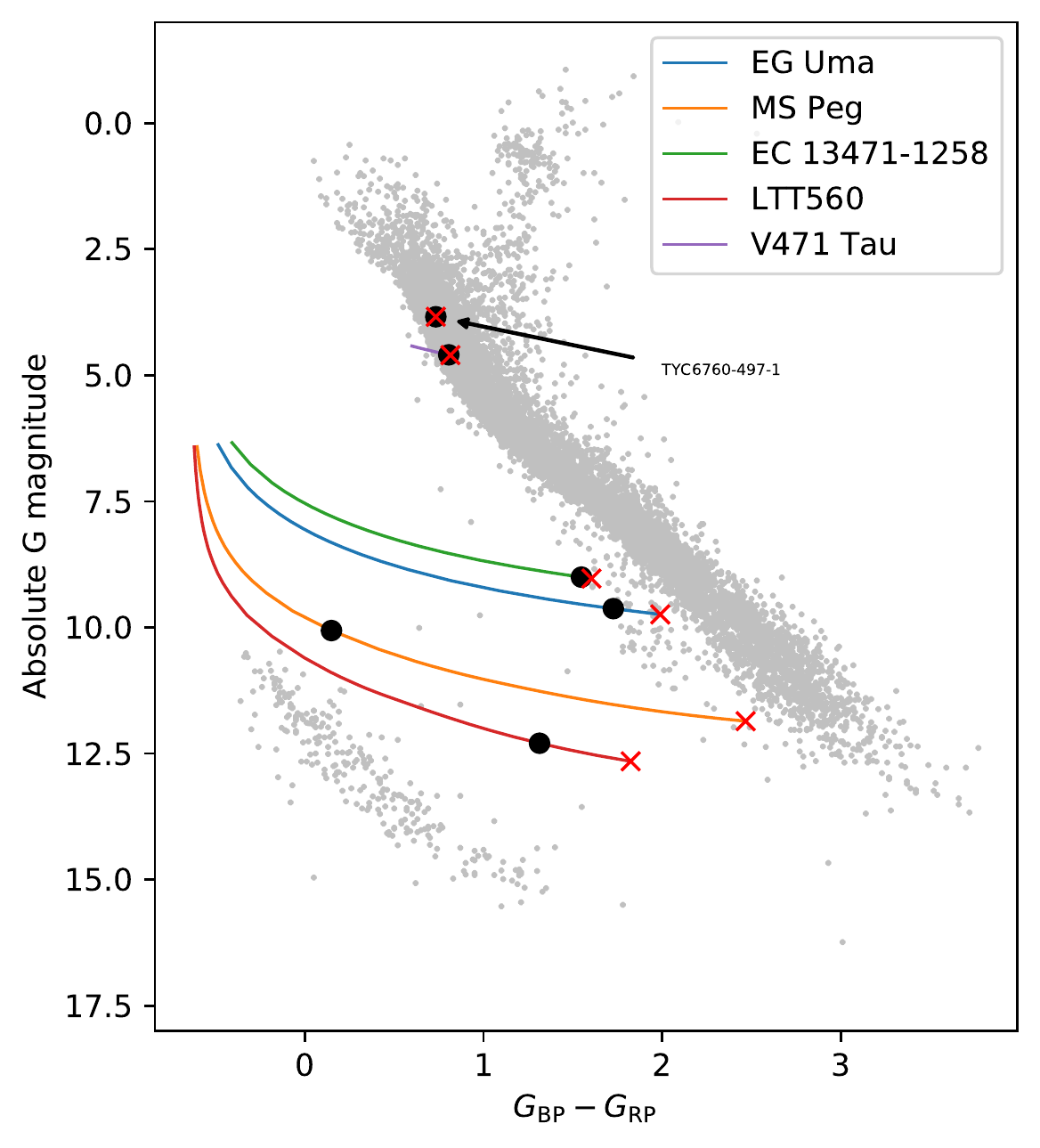}
\caption [.] {\label{fig:26}  
Left panel: The evolution of WD+M PCEBs across the \textit{Gaia} HR diagram is illustrated by the coloured tracks showing a range of companion spectral types, the black markers indicate a sequence of white dwarf temperatures. White dwarfs cool very fast initially and then gradually slow down (the cooling ages for a $0.6\,\mathrm{M}_{\sun}$ white dwarf for the five temperatures annotated in the left panel are $\simeq3.4\times10^5, 4.2\times10^6, 6.2\times10^7, 6.3\times10^8, 6.3\times10^9$\,yr)$^{8}$. This results in a selection effect as most (i.e. except the youngest) WD+M0 PCEBs will be seen near the main sequence whilst white dwarfs with less massive companions of the same age will be found in the area between the main sequence and the white dwarf cooling sequence. As the white dwarfs cool, and the flux contribution of the M-dwarfs remains constant, the systems fade and become redder, moving to the right in the HR diagram, until eventually merging into the main sequence. The WD+M Gold Sample is shown as red dots.  Right panel:  Evolutionary paths for five PCEBs adapted from \citep{2003A&A...406..305S} and \citep{2011A&A...536A..42Z}. Their current locations in the HR diagram are indicated by the black dot, and the positions where they will evolve into CVs once the companion star becomes Roche-lobe filling are shown by the red crosses. 

TYC~6760-497-1 \citep{2015MNRAS.452.1754P} is a WD+F type. Due to its high mass ratio, this system will undergo a short phase of thermal-timescale mass transfer during which the white dwarf will accrete at a rate that is sufficiently high to sustain nuclear shell burning \citep{1982ApJ...257..767F, 1982ApJ...259..244I}~--~and will be a super-soft X-ray source, and correspondingly bright and blue \citep{1997ARA&A..35...69K}.
} 
\end{figure*} 

\subsection{PCEBs}\label{sec:pcebevo}
A PCEB emerges from the common envelope as a detached binary consisting of a hot ($\simeq100\,000$\,K) white dwarf and a main sequence star companion with a spectral type A to M (and a small number of brown dwarfs).  These systems have typical orbital periods of hours to days,  are hence always spatially unresolved, and \textit{Gaia} will consequently detect and characterise them based on the combined flux from both stars. In the optical passbands of \textit{Gaia}, PCEBs with K-type or earlier companions are indistinguishable from single main-sequence stars even in the early stages of their evolution (Sect.\,\ref{sec:goldwdafgk}). 

As angular momentum losses drive the evolution of these binaries towards shorter orbital periods, the white dwarf gradually cools, fades, and becomes redder in colour~--~a process that takes $\simeq415$\,Myr to $2.6$\,Gyr to reach 10\,000K for white dwarfs with masses of  $0.4-1.2\,\mathrm{M}_{\sun}$. In contrast, companion stars with masses $\la0.9\,\mathrm{M}_{\sun}$ will not significantly evolve, and their flux contribution consequently remains constant throughout the evolution of the binary. 

We have computed the evolutionary tracks of a range of WD+M binaries by combining the synthetic $G$, $G_\mathrm{BP}$ and $G_\mathrm{RP}$ magnitudes for a typical $0.6\,\mathrm{M_{\sun}}$ white dwarf \citep{2011ApJ...730..128T,2011ApJ...737...28B}, with the cooling models provided by Pierre Bergeron\footnote{\href{http://www.astro.umontreal.ca/~bergeron/CoolingModels}{http://www.astro.umontreal.ca/$\sim$bergeron/CoolingModels} \citep{2006AJ....132.1221H, 2006ApJ...651L.137K, 2001PASP..113..409F, 2011ApJ...730..128T, 2020ApJ...901...93B}.}. The M-dwarf companions  were modelled using the Phoenix \citep{2013A&A...553A...6H} and ATLAS9 \citep{2003IAUS..210P.A20C} grids.

\begin{figure*} 
\centering
\includegraphics[width=17cm]{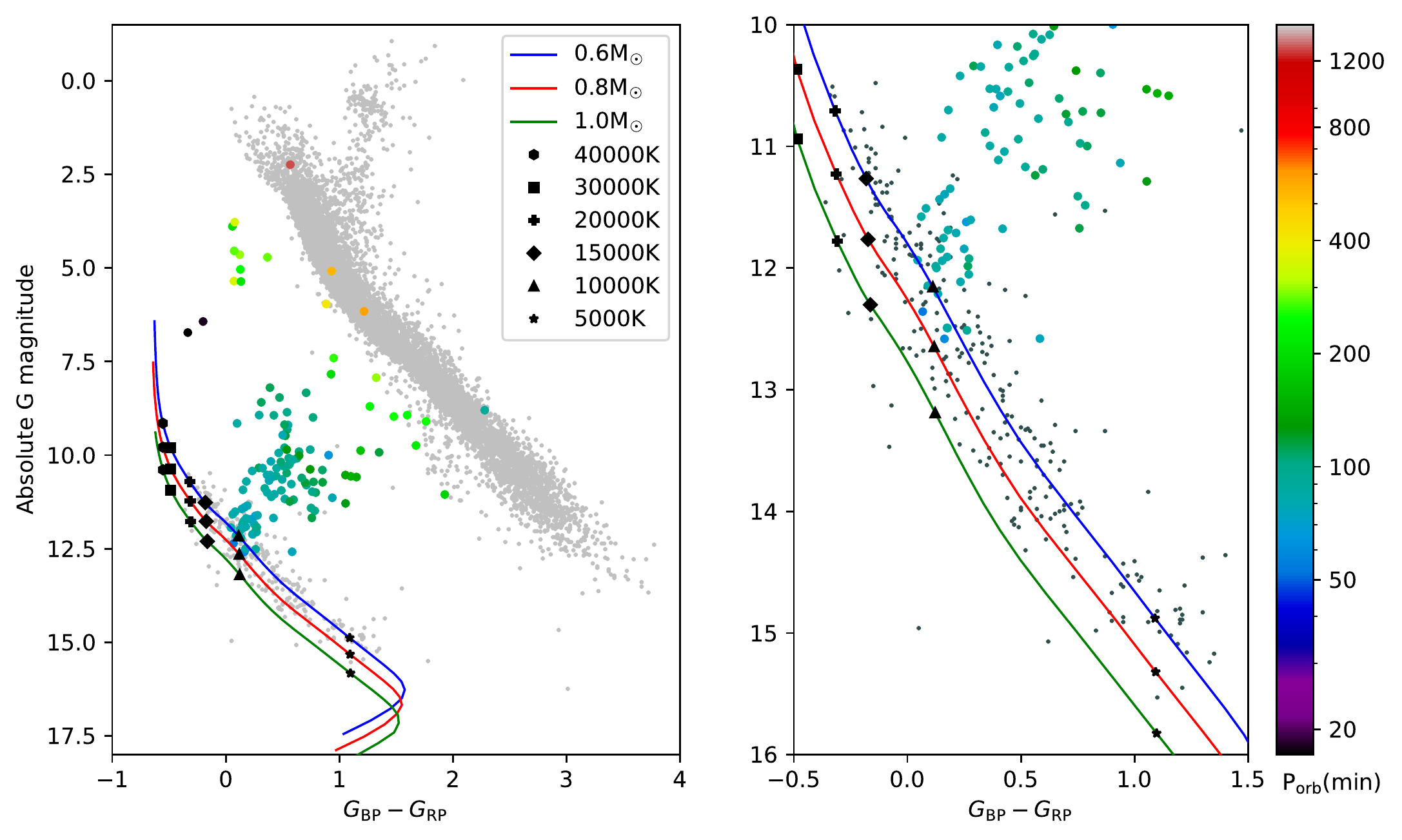}
\caption []{\label{fig:34}   The evolutionary channel from the main sequence is clearly evident with a wide range of long period CVs near the main sequence funnelling in as their periods fall towards a small locus in the white dwarf area of the HR diagram (left panel with expanded view in the right panel). By this time CVs have become dominated by their white dwarfs, following prolonged mass transfer. The effective temperature of the CV is essentially that of the white dwarf and is focused towards a region with  $T_\mathrm{eff}\simeq10\,000-15\,000$\,K. Subsequent evolution follows the cooling track of a white dwarf - the cooling tracks of the white dwarf component are shown for white dwarf masses of 0.6, 0.8, 1.0\,M$_{\sun}$ \citep{2011ApJ...737...28B}.} 
\end{figure*}

Comparison between these evolutionary tracks and the WD+M Gold Sample (Fig.\,\ref{fig:26}, left panel) shows reasonably good agreement, highlighting that WD+M PCEBs containing early or late M-type companions are relatively rare (see also Fig.\,9 in \citealt{2010MNRAS.402..620R}). This is likely a combination of the slope of the initial mass function of low-mass stars, which drops off for very low stellar masses  \citep{1993MNRAS.262..545K}\footnote{\citet{2005ApJS..161..394F} already noted (see their fig.\,6 \& 7) the very similar distribution in M-dwarf spectral types among a 20\,pc volume-limited sample of field M-dwarfs and their sample of WD+M binaries.} and the selection effects discussed in Sect.\,\ref{subsec:WD+m}. Similarly, WD+M PCEBs with very hot white dwarfs are relatively rare in the Gold Sample (again, see Fig.\,9 in  \citealt{2010MNRAS.402..620R}) which is a consequence of the fact that white dwarfs cool quasi-exponentially together with the selection effects examined in Sect.\,\ref{subsec:WD+m}.

Once the orbital separation reduces sufficiently for the companion to fill its Roche lobe, the system will morph into a CV, and the onset of mass transfer will drastically alter its properties within the HR diagram (Sect.\,\ref{sec:CVs} and \ref{sec:cvevo}). We have used the parameters of five PCEBs in the WD+M and WD+AFGK Gold Samples to compute their tracks within the HR diagram, illustrating both their past evolution as well as the location at which they will transition into CVs and hence disappear from the PCEB population (Fig.\,\ref{fig:26}, right panel). Except for LTT\,560, which contains a relatively cool M-dwarf, the onset of mass transfer will occur close to the main sequence, when the white dwarf has cooled below 10\,000\,K \citep{2003A&A...406..305S, 2010A&A...520A..86Z}. V471\,Tau illustrates that PCEBs with a K-type companion exit the common envelope very close to the main sequence, and rapidly descend into it (even though the white dwarf in this system is still fairly hot, $\simeq35\,000$\,K, \citealt{2001ApJ...563..971O}).

Although detached white dwarf plus main-sequence binaries are mostly formed via common envelope evolution, the standard theory of CV evolution predicts that some WD+M binaries with periods of $\simeq2-3$\,h are actually CVs in which the donor detached from its Roche lobe \citep{1983ApJ...275..713R, 2008MNRAS.389.1563D, 2016MNRAS.457.3867Z}. Angular momentum loss will still drive the evolution of these systems towards shorter orbital periods until the donor fills its Roche lobe again at $P_\mathrm{orb}\simeq2$\,h, when they will transit back into the CV population. \citet{2020MNRAS.492L..40A} have suggested that these ``interlopers'' among the WD+M PCEBs population may be identified because of their, on average, higher white dwarf mass \citep{2011A&A...536A..42Z}, resulting in larger absolute magnitudes compared to the WD+M PCEBs that have not yet reached the first onset of mass transfer.

A final note concerns those white dwarfs with companions that are sufficiently massive ($\ga0.9\,\mathrm{M_{\sun}}$) to evolve off the main sequence within a Hubble time. In the \textit{Gaia} HR diagram, those systems will emerge from the common envelope within the main-sequence. Given their relatively small orbital separations, they will initiate mass transfer before the companion evolves significantly into the giant branch, and will hence transform into CVs ``in situ''. An interesting example is the WD+F binary TYC~6760-497-1 \citep{2015MNRAS.452.1754P}: because of the high mass ratio, $M_2/M_\mathrm{wd}\simeq2$, this system will undergo a short phase of thermal-timescale mass transfer during which the white dwarf will accrete at a rate that is sufficiently high to sustain nuclear shell burning \citep{1982ApJ...257..767F, 1982ApJ...259..244I}~--~and will be a super-soft X-ray source, and correspondingly bright and blue \citep{1997ARA&A..35...69K}. There is no super-soft source known within the 300\,pc CV Gold Sample.

\subsection{CVs}
\label{sec:cvevo}
CVs commence mass transfer resulting in the formation of a luminous accretion flow, and heating of the surface of the white dwarf \citep{2003ApJ...596L.227T}. As discussed in Sect.\,\ref{sec:CVs}, the location of a CV within the HR diagram depends on the relative contributions of the white dwarf, the donor star, and the accretion flow to the overall flux.

As shown in the previous section, the location of most CVs within the HR diagram at the onset of mass transfer is expected to be close to, or within, the main sequence. Once a companion in a PCEB fills its Roche lobe, the system will very quickly move blue-wards, and, if the mass transfer rate is within the regime of dwarf novae, sit slightly below the main sequence (Fig.\,\ref{fig:29}). The distinct cluster of novalike CVs (see Figure \ref{fig:29}) with bright, blue accretion discs and very similar orbital periods of $\simeq3-4$\,h, i.e. sandwiched between the longer-period U\,Gem type dwarf novae, and the shorter-period SU\,UMa type dwarf novae, is noteworthy: these systems appear to have mass transfer rates that exceed the predictions of the standard evolution model \citep{2009ApJ...693.1007T}, and very likely reflect a peculiar phase in CV evolution, and/or a specific set of initial parameters. 

\begin{figure*} 
\centering
\includegraphics[width=17cm]{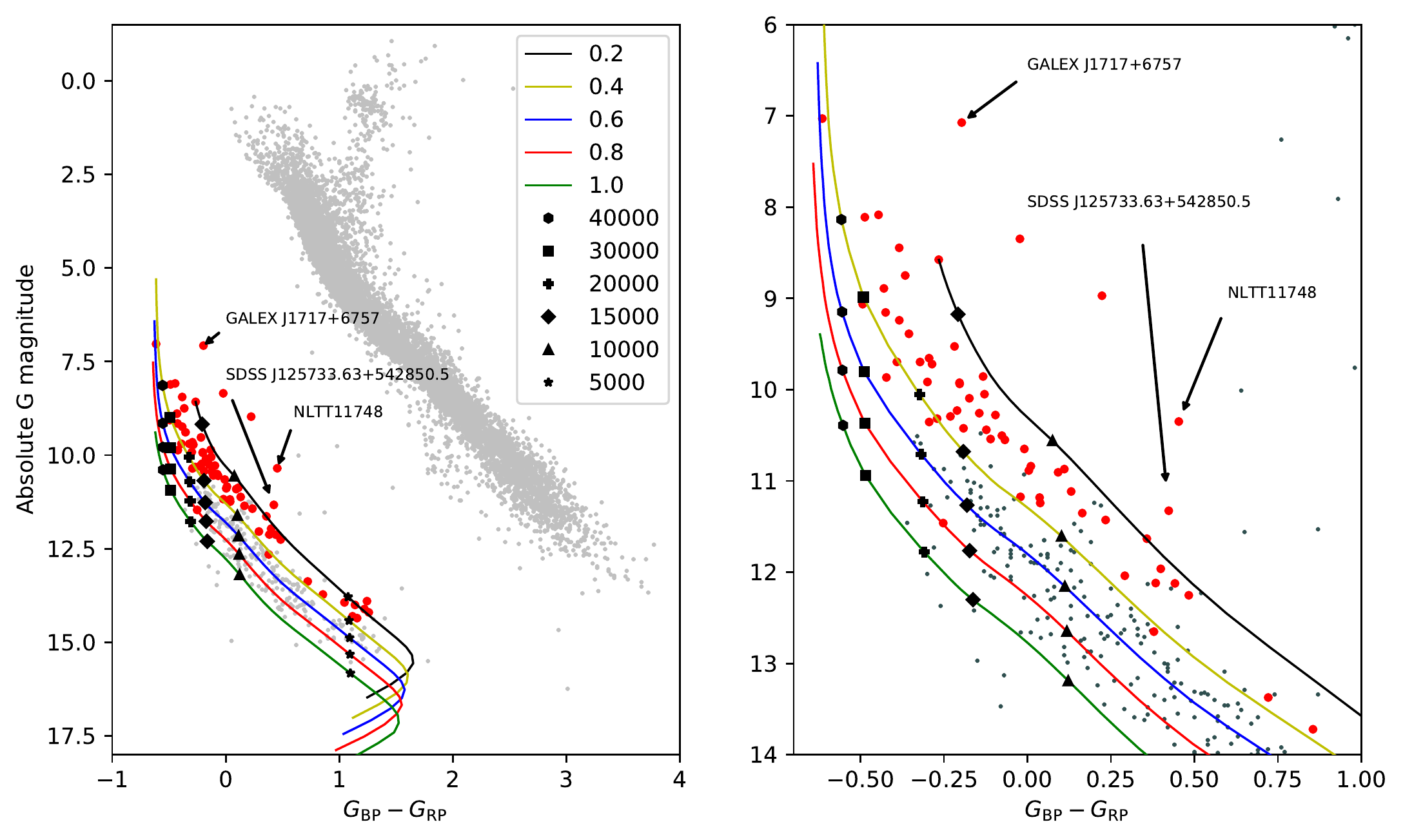}
\caption []{\label{fig:35}   The position of the DWD Gold Sample on the HR diagram together with the cooling tracks of single white dwarfs for masses of 0.2, 0.4, 0.6, 0.8, 1.0 $M_{\sun}$ \citep{2011ApJ...737...28B}. The colours of the DWDs will be the weighted average of the colours of its constituents whilst the absolute magnitude will be the sum of their fluxes. The position will therefore be similar to that of a single white dwarf but $\simeq0.75$ magnitudes brighter~--~in practice the more massive white dwarf will have formed first and hence have progressed further down the cooling track leaving the secondary dominant. The clumping between the $0.2\,\mathrm{M}_{\sun}$ and $0.4\,\mathrm{M}_{\sun}$ cooling curves indicates that the brighter star in these systems typically has a low mass ($\lesssim0.4\,\mathrm{M}_{\sun}$), highlighting possible observational selection effects. \textit{GALEX} J1717+6757 and NLTT11748 are ELMs with $M_\mathrm{wd}\lesssim0.2\,\mathrm{M}_{\sun}$, which, despite containing relatively cool white dwarfs ($\simeq10\,000-15\,000$\,K) are substantially more luminous thanks to their large radii. \citep{2010A&A...516L...7K, 2010ApJ...716L.146S, 2011ApJ...737L..16V, 2014MNRAS.444.1674H}
}  
\end{figure*}

As CVs evolve the mass, radius, and effective temperature of the secondary decrease, and so does the mass transfer rate, and hence the  evolutionary curve moves back towards the white dwarf cooling sequence. This explains the funnel-shaped distribution of CVs in  Fig.\,\ref{fig:16} (see also \citealt{2020MNRAS.492L..40A}), which clearly illustrates the decrease of $P_\mathrm{orb}$ as the systems come closer to the white dwarf cooling sequence. An interesting question is whether the $2-3$\,h period gap, i.e. when CV donors detach from their Roche lobes and the systems turn into detached WD+M binaries (see previous section) will leave a detectable imprint in the population of CVs within the \textit{Gaia} HR diagram. We would argue that, at the moment, the number of CVs within a well-defined sample with accurate \textit{Gaia} data is not sufficiently large to assess this question, and we caution that the intrinsic variability of CVs might smear out the effect of the period gap within the HR diagram.

As the mass of the donors continues to decrease, they will eventually change into brown dwarfs\footnote{The CV Gold Sample contains at least five potentially ``period bouncer'' systems with brown dwarf donors: 
1RXS\,J105010.3-140431 \citep{2001A&A...376..448M}, GD\,552  \citep{2008MNRAS.388..889U}, SDSS\,J102905.21+485515.2 \citep{2016AJ....152..226T}, SDSS\,J143317.78+101123.3 \citep{2016Natur.533..366H}), QZ\,Lib, \citep{2018MNRAS.481.2523P}.}. These are inherently faint systems, and their optical appearance is entirely dominated by the accretion-heated white dwarf, hence their location lies very close to or within the white dwarf cooling track. It is noticeable that the evolutionary path of CVs ends in a small area of the HR diagram corresponding to white dwarf temperatures of $\simeq10\,000$\,K (Fig.\,\ref{fig:34}, right panel). The mass transfer rates implied by these accretion rates, $\dot M\simeq5\times10^{-11}\,\mathrm{M_{\sun}\,yr^{-1}}$ are consistent with the predictions for systems that have reached the period minimum, however, as CVs evolve back towards longer periods \citep{2001ApJ...550..897H}, the mass transfer rates are predicted to drop by an order of magnitude~--~the known ``period bouncers'' have mass accretion rates much higher than these predictions. This exhibits the problem of the ``missing period bouncers'' \citep{2001ApJ...550..897H, 2020MNRAS.494.3799P}. 

A final note concerns the evolution of AM\,CVn stars within the \textit{Gaia} HR diagram. As they may be formed via three different channels, their locations at the onset of mass transfer may radically differ. In the case of a double-degenerate progenitor \citep{1967AcA....17..287P, 2007MNRAS.381..525D}, they will start mass transfer in, or close to, the white dwarf cooling sequence. In the case of a white dwarf plus a helium-burning star \citep{1986A&A...155...51S, 2008AstL...34..620Y}, they would probably start mass transfer within the region occupied by subdwarfs ($M_G\simeq4$, $G_\mathrm{BP}-G_\mathrm{RP}\simeq-0.5$, \citealt{2020A&A...635A.193G}) . Finally, if they descend from CVs with nuclearly evolved donor stars \citep{2003MNRAS.340.1214P, 2015ApJ...809...80G}, they would begin mass transfer within the main sequence, and TYC~6760-497-1 \citep{2015MNRAS.452.1754P} may be a relevant example (Fig.\,\ref{fig:26}). The AM\,CVn stars within the Gold Sample appear bi-modal, with two luminous blue ``high state'' systems , and the remaining four ``low state'' systems close to the white dwarf cooling sequence. This may be a result of small number statistics, the larger sample in fig.\,2 of \citet{2018A&A...620A.141R} exhibits a larger spread~--~however, that sample is very likely subject to observational biases. The two luminous AM\,CVn stars within the Gold Sample (AM\,CVn with a period of 17.14\,min and HP\,Lib with a period of 18.38\,min) have high mass transfer rates that maintain their accretion discs in a hot, stable state~--~in other words, they are miniature novalike variables. They appear bluer and less luminous than the ``normal'' novalikes with typical periods of $3-4$\,h,  reflecting their much more compact size with their spectral energy distribution (SED) dominated by very small and hot accretion discs.

\subsection {DWDs}
Double white dwarfs tend to sit towards, or just above, the bright edge of the white dwarf cooling sequence. Just as single white dwarfs, they will evolve along the cooling track, though not necessarily follow any single-mass track, as the relative contribution of the two stars gradually changes (Figure \ref{fig:35}). With the exception of the extremely low mass white dwarfs mentioned earlier (well above even the $0.2\,\mathrm{M}_{\sun}$ cooling track) the majority of the Gold Sample sits around the cooling curves of $0.2-0.4\,\mathrm{M}_{\sun}$ single white dwarfs. This implies that the brighter dwarf is of low mass~--~the companion may be an older, and cooler, higher mass white dwarf (see \citealt{2020A&A...638A.131N}). This almost certainly represents a selection effect, as the brighter low-mass white dwarfs in these systems are easier two find. There are a few DWDs scattered within the cooling curves of higher mass white dwarfs, which suggests that the Gold Sample captures the range of mass ratios found among DWDs, though may not be representative. An exception is SDSS\,J125733.63+542850.5 

\citep{2015MNRAS.450.3966B} (see Fig.\, \ref{fig:35}) which paradoxically has a low mass white dwarf that is older ($\simeq5$\,Gyr) than the more massive one ($\sim1$\,Gyr), highlighting the potentially complex intricacies of the formation of DWDs. As noted in Sect.\,\ref{sec:dwd_hrd}, the Gold Sample has a sharp cut-off at $G_\mathrm{BP}-G_\mathrm{RP}\simeq0.6$, corresponding to $\simeq7000\,K$ due to the disappearance of the Balmer lines.

\section{Conclusions}
We have defined ``Gold'' reference samples within $d<300$\,pc of the most common populations of close white dwarf binaries: CVs, PCEBs and DWDs. We have used those samples, which encompass most of the parameter space of physical properties (component masses, orbital periods, ages) spanned by close white dwarf binaries, to map out their distribution in the \textit{Gaia} HR diagram. CVs are inherently easier to discover than PCEBs and DWDs. Whereas the CV Gold Sample is likely to be $\simeq50$\,per cent complete, this figure drops to $\sim10$\,per cent for PCEB (with a large uncertainty) and less than 0.2\,per cent for DWDs. We have furthermore explored how close white dwarf binaries evolve through the \textit{Gaia} HR diagram, criss-crossing between the white dwarf cooling track and the main sequence,  occupying similar parameter spaces during different evolutionary phases. We have investigated the differences between magnitude and volume limited close white dwarf binary samples, and have demonstrated the importance of a volume-limited sample to further improve our understanding of the evolution of these systems. We will use the Gold Samples established here to develop selection functions that will reliably identify close white dwarf binary candidates for spectroscopic and photometric confirmation, with the overarching goal to assemble a volumetric 300\,pc sample that is representative of the underlying population of these systems, and sufficiently large to provide accurate statistical insight into the relative numbers, and physical properties of the various subtypes.

\section*{Acknowledgements}
BTG and TRM were supported by the UK STFC grant ST/T000406/1 and by a Leverhulme Research Fellowship. BTG also acknowledges the support of a Leverhulme Research Fellowship. RR has received funding from the postdoctoral fellowship programme Beatriu de Pin\'os, funded by the Secretary of Universities and Research (Government of Catalonia) and by the Horizon 2020 programme of research and innovation of the European Union under the Maria Sk\l{}odowska-Curie grant agreement No 801370.
EB was supported by UK Research and Innovation grant ST/S000623/1.
Based on observations made with ESO Telescopes at the La Silla Paranal Observatory under programme ID 094.D-0344

This work has made use of data from the European Space Agency (ESA) mission {\it Gaia} (\url{https://www.cosmos.esa.int/gaia}), processed by the {\it Gaia} Data Processing and Analysis Consortium (DPAC, \url{https://www.cosmos.esa.int/web/Gaia/dpac/consortium}). Funding for the DPAC has been provided by national institutions, in particular the institutions participating in the {\it Gaia} Multilateral Agreement. We acknowledge ESA Gaia, DPAC and the Photometric Science Alerts Team (http://gsaweb.ast.cam.ac.uk/alerts)

Funding for the Sloan Digital Sky Survey IV has been provided by the Alfred P. Sloan Foundation, the U.S. Department of Energy Office of Science, and the Participating Institutions. SDSS-IV acknowledges support and resources from the Center for High-Performance Computing at the University of Utah. The SDSS web site is \url{www.sdss.org}.

SDSS-IV is managed by the Astrophysical Research Consortium for the  Participating Institutions of the SDSS Collaboration including the  Brazilian Participation Group, the Carnegie Institution for Science,  Carnegie Mellon University, the Chilean Participation Group, the French Participation Group, Harvard-Smithsonian Center for Astrophysics,  Instituto de Astrof\'isica de Canarias, The Johns Hopkins University, Kavli Institute for the Physics and Mathematics of the Universe (IPMU) / University of Tokyo, the Korean Participation Group, Lawrence Berkeley National Laboratory, Leibniz Institut f\"ur Astrophysik Potsdam (AIP), Max-Planck-Institut f\"ur Astronomie (MPIA Heidelberg), Max-Planck-Institut f\"ur Astrophysik (MPA Garching), Max-Planck-Institut f\"ur Extraterrestrische Physik (MPE),  National Astronomical Observatories of China, New Mexico State University, New York University, University of Notre Dame, Observat\'ario Nacional / MCTI, The Ohio State University, Pennsylvania State University, Shanghai Astronomical Observatory, United Kingdom Participation Group,Universidad Nacional Aut\'onoma de M\'exico, University of Arizona, University of Colorado Boulder, University of Oxford, University of Portsmouth, University of Utah, University of Virginia, University of Washington, University of Wisconsin, Vanderbilt University, and Yale University.

The CSS survey is funded by the National Aeronautics and Space Administration under Grant No. NNG05GF22G issued through the Science Mission Directorate Near-Earth Objects Observations Program.  The CRTS survey is supported by the U.S.~National Science Foundation under grants AST-0909182.

Guoshoujing Telescope (the Large Sky Area Multi-Object Fiber Spectroscopic Telescope LAMOST) is a National Major Scientific Project built by the Chinese Academy of Sciences. Funding for the project has been provided by the National Development and Reform Commission. LAMOST is operated and managed by the National Astronomical Observatories, Chinese Academy of Sciences. 

Based on observations obtained with the Samuel Oschin 48-inch Telescope at the Palomar Observatory as part of the Zwicky Transient Facility project. ZTF is supported by the National Science Foundation under Grant No. AST-1440341 and a collaboration including Caltech, IPAC, the Weizmann Institute for Science, the Oskar Klein Center at Stockholm University, the University of Maryland, the University of Washington, Deutsches Elektronen-Synchrotron and Humboldt University, Los Alamos National Laboratories, the TANGO Consortium of Taiwan, the University of Wisconsin at Milwaukee, and Lawrence Berkeley National Laboratories. Operations are conducted by COO, IPAC, and UW.
\section*{Data Availability}
The data underlying this article are available from the sources referenced in the text. The results of our analysis are provided  and in the online supplementary material.



\bibliographystyle{mnras}
\bibliography{refs} 


\clearpage
\appendix
\section{The Gold samples}
We provide the Gold samples in the Supplementary Information as data files in CSV format. Missing information for individual properties of some systems are left as blank entries in these CSV files.

\begin{table*}
\caption{The WD+M Gold Sample.\label{tab:app_WD+M}}
\begin{tabular}{lll}
\hline
\multicolumn{1}{|l|}{Field}       & \multicolumn{1}{l|}{Units} & \multicolumn{1}{l|}{Description}                                        \\ \hline                                                            
name           &            & Identifier                                           \\
source\_id          &            & \textit{Gaia}\,DR2 unique source identifier          \\
references          &            & References to literature provided as ADS bibcodes      \\
ra                  & deg        & Barycentric right ascension (ICRS) at Epoch 2015.5   \\
dec                 & deg        & Barycentric declination (ICRS) at Epoch 2015.5       \\
type                &            & Type of white dwarf and companion where known (PREP=pre-polar,BD=brown dwarf) \\
teffwd              & K          & White dwarf effective temperature where known (blank otherwise)         \\
logg                & log (cm~s$^{-2}$)     & Stellar log(surface gravity)  where known (blank otherwise)  \\
porb                & hours      & Orbital period where known (blank otherwise)                             \\
sp                  &            & Spectral type of (M-type) companion (0-9) where known, blank or -1 otherwise  \\
parallax            & mas        & Absolute stellar parallax                                               \\
parallax\_error     & mas        & Uncertainty in parallax                                              \\
pmra                & mas/year   & Proper motion in right ascension direction (pmRA*cosDE)                 \\
pmdec               & mas/year   & Proper motion in declination direction                                  \\
phot\_g\_mean\_mag  &            & $G$-band mean magnitude (Vega)                                            \\
phot\_g\_mean\_mag\_error&           & Uncertainty in G-band mean magnitude                                 \\
phot\_bp\_mean\_mag &            & Integrated $G_\mathrm{BP}$ mean magnitude (Vega)                                     \\
phot\_bp\_mean\_mag\_error&      & Uncertainty in $G_\mathrm{BP}$-band mean magnitude                                 \\
phot\_rp\_mean\_mag &            & Integrated $G_\mathrm{RP}$ mean magnitude (Vega)                                     \\
phot\_rp\_mean\_mag\_error&      & Uncertainty in $G_\mathrm{RP}$-band mean magnitude                                 \\
l                   & deg        & Galactic longitude                                                      \\
b                   & deg        & Galactic latitude                                                       \\
ruwe                &            & \textit{Gaia} renormalised unit weight error                                     \\
r\_est              & parsec     & Estimated distance                                                      \\
r\_est\_error        & parsec     & Uncertainty in estimated distance                                    \\
galex\_coverage     &            & 1 if this location is present in one or more \textit{GALEX} tiles, 0   otherwise \\
objid               &            & \textit{GALEX} identifier for the source                                         \\
fuv\_mag            &            & \textit{GALEX} $FUV$ calibrated magnitude in AB system                             \\
fuve                &            & Uncertainty in \textit{GALEX} $FUV$                                              \\
nuv\_mag            &            & \textit{GALEX} $NUV$ calibrated magnitude in AB system                            \\
nuve                &            & Uncertainty in \textit{GALEX} $NUV$   \\ \hline                  
\end{tabular}
\end{table*}


\begin{table*}
\caption{The WD+AFGK Gold Sample:\label{tab:app_WD+AFGK}}
\begin{tabular}{lll}
\hline
\multicolumn{1}{|l|}{Field}       & \multicolumn{1}{l|}{Units} & \multicolumn{1}{l|}{Description}                                        \\ \hline
name                              &                            & Name
                    \\
source\_id                        &                            & Gaia DR2 unique source identifier                                       \\
references                            &                            & References to literature
                     \\
ra                                & deg                        & Barycentric right ascension (ICRS) at Ep=2015.5                         \\
dec                               & deg                        & Barycentric declination (ICRS) at Ep=2015.5                             \\
teff                              & K                          & Stellar effective temperature where known (blank otherwise)          \\
porb                        & hours                       & Orbital period where known (blank otherwise)
                    \\
parallax                          & mas                        & Absolute stellar parallax                                               \\
parallax\_error                   & mas                        & Uncertainty in parallax                                              \\
pmra                              & mas/year                   & Proper motion in right ascension direction (pmRA$\times$cosDE)                 \\
pmdec                             & mas/year                   & Proper motion in declination direction                                  \\
phot\_g\_mean\_mag  &            & G-band mean magnitude (Vega)                                            \\
phot\_g\_mean\_mag\_error&           & Uncertainty in G-band mean magnitude                                 \\
phot\_bp\_mean\_mag &            & Integrated BP mean magnitude (Vega)                                     \\
phot\_bp\_mean\_mag\_error&          & Uncertainty in BP-band mean magnitude                                 \\
phot\_rp\_mean\_mag &            & Integrated RP mean magnitude (Vega)                                     \\
phot\_rp\_mean\_mag\_error&           & Uncertainty in RP-band mean magnitude                                 \\                            
l                                 & deg                        & Galactic longitude                                                      \\
b                                 & deg                        & Galactic latitude                                                       \\
ruwe                              &                            & Gaia renormalised unit weight error                                     \\
r\_est                            & parsec                     & Estimated distance                                                      \\
r\_est\_error        & parsec     & Uncertainty in estimated distance                                    \\
galex\_coverage                   &                            & 1 if this location is present in one or more \textit{GALEX} tiles, 0   otherwise \\
objid                             &                            & \textit{GALEX} identifier for the source                                         \\
fuv\_mag            &            & \textit{GALEX} FUV calibrated magnitude in AB system                             \\
fuv\_magerr                &            & Uncertainty in \textit{GALEX} FUV                                              \\
nuv\_mag            &            & \textit{GALEX} NUV calibrated magnitude in AB system                            \\
nuv\_magerr                &            & Uncertainty in \textit{GALEX} NUV     \\ \hline
\end{tabular}
\end{table*}


\begin{table*}
\caption{CV Gold Sample:\label{tab:app_CV}}
\begin{tabular}{lll}
\hline
\multicolumn{1}{|l|}{Field}    & \multicolumn{1}{l|}{Units} & \multicolumn{1}{l|}{Description}                                        \\ \hline
name                           &                            & Identifier                                                              \\
source\_id                     &                            & Gaia DR2 unique source identifier                                       \\
references                            &                         & References to literature
                     \\
ra                             & deg                        & Barycentric right ascension (ICRS) at Ep=2015.5                         \\
dec                            & deg                        & Barycentric declination (ICRS) at Ep=2015.5                             \\
cv\_type                       &                            & CV type (Novalike, AM\,CVn, WZ\,Sge, Magnetic, U Gem or SU\,UMa) or CV if type is unknown          \\

porb                      & hours                       & Orbital period where known (blank otherwise)                                                         \\
parallax                       & mas                        & Absolute stellar parallax                                               \\
parallax\_error                & mas                        & Uncertainty in parallax                                              \\
pmra                           & mas/year                   & Proper motion in right ascension direction (pmRA*cosDE)                 \\
pmdec                          & mas/year                   & Proper motion in declination direction                                  \\
phot\_g\_mean\_mag  &            & G-band mean magnitude (Vega)                                            \\
phot\_g\_mean\_mag\_error&           & Uncertainty in G-band mean magnitude                                 \\
phot\_bp\_mean\_mag &            & Integrated BP mean magnitude (Vega)                                     \\
phot\_bp\_mean\_mag\_error&          & Uncertainty in BP-band mean magnitude                                 \\
phot\_rp\_mean\_mag &            & Integrated RP mean magnitude (Vega)                                     \\
phot\_rp\_mean\_mag\_error&           & Uncertainty in RP-band mean magnitude                                 \\
l                              & deg                        & Galactic longitude                                                      \\
b                              & deg                        & Galactic latitude                                                       \\
ruwe                           &                            & Gaia renormalised unit weight error                                     \\
r\_est                         & parsec                     & Estimated distance                                                      \\
r\_est\_error        & parsec     & Uncertainty in estimated distance                                    \\
galex\_coverage          &                            & 1 if this location is present in one or more \textit{GALEX} tiles, 0   otherwise \\
objid                  &                            & \textit{GALEX} identifier for the source                                         \\
fuv\_mag            &            & \textit{GALEX} FUV calibrated magnitude in AB system                             \\
fuv\_magerr                &            & Uncertainty in \textit{GALEX} FUV                                              \\
nuv\_mag            &            & \textit{GALEX} NUV calibrated magnitude in AB system                            \\
nuv\_magerr                &            & Uncertainty in \textit{GALEX} NUV     \\
\hline
\end{tabular}
\end{table*}


\begin{table*}
\caption{Confirmed CVs with $d<300$\,pc excluded from the Gold sample because of data quality issues: \label{tab:app_CV_failed}}
\begin{tabular}{lll}
\hline
\multicolumn{1}{|l|}{Field}    & \multicolumn{1}{l|}{Units} & \multicolumn{1}{l|}{Description}                                        \\ \hline
name                           &                            & Identifier                                                              \\
source\_id                     &                            & Gaia DR2 unique source identifier                                       \\
references                            &                         & References to literature
                     \\
ra                             & deg                        & Barycentric right ascension (ICRS) at Ep=2015.5                         \\
dec                            & deg                        & Barycentric declination (ICRS) at Ep=2015.5                             \\
cv\_type                       &                            & CV type (Novalike, AM CVn, WZ Sge, Magnetic, U Gem or SU Uma) or CV if type is unknown             \\

porb                      & hours                       & Orbital period where known (0 otherwise)                                                         \\
parallax                       & mas                        & Absolute stellar parallax                                               \\
parallax\_error                & mas                        & Uncertainty in parallax                                              \\
pmra                           & mas/year                   & Proper motion in right ascension direction (pmRA*cosDE)                 \\
pmdec                          & mas/year                   & Proper motion in declination direction                                  \\
phot\_g\_mean\_mag  &            & G-band mean magnitude (Vega)                                            \\
phot\_g\_mean\_mag\_error&           & Uncertainty in G-band mean magnitude                                 \\
phot\_bp\_mean\_mag &            & Integrated BP mean magnitude (Vega)                                     \\
phot\_bp\_mean\_mag\_error&          & Uncertainty in BP-band mean magnitude                                 \\
phot\_rp\_mean\_mag &            & Integrated RP mean magnitude (Vega)                                     \\
phot\_rp\_mean\_mag\_error&           & Uncertainty in RP-band mean magnitude                                 \\
l                              & deg                        & Galactic longitude                                                      \\
b                              & deg                        & Galactic latitude                                                       \\
ruwe                           &                            & Gaia renormalised unit weight error                                     \\
r\_est                         & parsec                     & Estimated distance                                                      \\
r\_est\_error        & parsec     & Uncertainty in estimated distance                                    \\
galex\_coverage           &                            & 1 if this location is present in one or more \textit{GALEX} tiles, 0   otherwise \\
objid                  &                            & \textit{GALEX} identifier for the source                                         \\
fuv\_mag            &            & \textit{GALEX} FUV calibrated magnitude in AB system                             \\
fuv\_magerr                &            & Uncertainty in \textit{GALEX} FUV                                              \\
nuv\_mag            &            & \textit{GALEX} NUV calibrated magnitude in AB system                            \\
nuv\_magerr                &            & Uncertainty in \textit{GALEX} NUV     \\
\hline
\end{tabular}
\end{table*}
\clearpage

\begin{table*} 
\caption {
The DWD Gold Sample: \label{tab:App_DWD}}
\begin{tabular}{lllllll}
\cline{1-3}
\multicolumn{1}{|l|}{Field}    & \multicolumn{1}{l|}{Units} & \multicolumn{1}{l|}{Description}                                        &  &  &  &  \\ \cline{1-3}
name                     &                            & Identifier             &  &  &  &  \\
source\_id                     &                            & Gaia DR2 unique source identifier                                       &  &  &  &  \\
references                     &                            & References to literature                                       &  &  &  &  \\ra                             & deg                        & Barycentric right ascension (ICRS) at Ep=2015.5                         &  &  &  &  \\
dec                            & deg                        & Barycentric declination (ICRS) at Ep=2015.5                             &  &  &  &  \\
porb                 & hours                       & Orbital period where known (blank otherwise)           &  &  &  &  \\
parallax                       & mas                        & Absolute stellar parallax                                               &  &  &  &  \\
parallax\_error                & mas                        & Uncertainty in parallax                                              &  &  &  &  \\
pmra                           & mas/year                   & Proper motion in right ascension direction (pmRA*cosDE)                 &  &  &  &  \\
pmdec                          & mas/year                   & Proper motion in declination direction                                  &  &  &  &  \\
phot\_g\_mean\_mag  &            & G-band mean magnitude (Vega)                                            \\
phot\_g\_mean\_mag\_error&           & Uncertainty in G-band mean magnitude                                 \\
phot\_bp\_mean\_mag &            & Integrated BP mean magnitude (Vega)                                     \\
phot\_bp\_mean\_mag\_error&          & Uncertainty in BP-band mean magnitude                                 \\
phot\_rp\_mean\_mag &            & Integrated RP mean magnitude (Vega)                                     \\
phot\_rp\_mean\_mag\_error&           & Uncertainty in RP-band mean magnitude                                 \\
l                              & deg                        & Galactic longitude                                                      &  &  &  &  \\
b                              & deg                        & Galactic latitude                                                       &  &  &  &  \\
ruwe                           &                            & Gaia   renormalised unit weight error                                   &  &  &  &  \\
r\_est                         & parsec                     & Estimated distance                                                      &  &  &  &  \\
r\_est\_error        & parsec     & Uncertainty in estimated distance                                    \\
galex\_coverage                &                            & 1 if this location is present in one or more \textit{GALEX} tiles, 0   otherwise &  &  &  &  \\
objid                          &                            & \textit{GALEX}   identifier for the source                                       &  &  &  &  \\
fuv\_mag            &            & \textit{GALEX} FUV calibrated magnitude in AB system                             \\
fuv\_magerr                &            & Uncertainty in \textit{GALEX} FUV                                              \\
nuv\_mag            &            & \textit{GALEX} NUV calibrated magnitude in AB system                            \\
nuv\_magerr                &            & Uncertainty in \textit{GALEX} NUV     \\
\hline
\end{tabular}
\end{table*}

\clearpage

\bsp	
\label{lastpage}
\end{document}